\documentclass[12pt,preprint]{aastex}

\shorttitle{M31 Supersoft and Quasi-Soft Sources observed with  XMM-Newton}
\shortauthors{Orio}

\begin{document}


\title{A Close Look at the Population
 of Supersoft and Quasi-Soft X-Ray Sources  Observed  in M31 with XMM-Newton
}


\author{Marina Orio}
\affil{INAF, Osservatorio Astronomico di Torino, Strada Osservatorio, 20,
I-10025 Pino Torinese (TO), Italy\\
and Department of Astronomy, 475 N. Charter Str. University of Wisconsin, 
Madison WI 53706}
\email{orio@astro.wisc.edu}



\begin{abstract}

 The deepest X-ray images of M31, obtained with  XMM-Newton, are examined 
 to derive spectral and statistical properties
of the population of the softest X-ray sources. 
 Classifying supersoft X-ray sources (SSS)
with  criteria based on  the same hardness ratios  
 defined for recent Chandra observations, a quarter
 of the selected SSS  turn out to be supernova remnants (SNR). Another quarter 
 of SSS are spatially coincident with recent classical novae
(but they are less than 10\% of the nova population observed
in the last 25 years).  
Only 3 among 15 non-SNR SSS show clear variability 
with X-ray flux variation of more than one order of magnitude
 within  few months. Two of these sources display additional, smaller
 amplitude variability on time scales of several minutes.   Their
 broad band spectra and those of the novae 
are approximately fit with a blackbody or white dwarf atmospheric
model at near-Eddington luminosity for the distance of 
M31.  Two SSS appear to reach 
very large, perhaps super-Eddington luminosities for part of the time,
 and probably eject
material in  a wind until the luminosity decreases again after 
a few  months. One of the two objects has some
 characteristics in common with Ultra Luminous
 X-ray Sources observed outside the Local Group.
 
Most Quasi-Soft Sources (QSS),  among which also a few SNR are 
selected using the hardness ratio criteria, 
 are repeatedly detected. Several QSS are better fit by
 a power law spectrum, but some faint, apparently blackbody-like QSS
 with temperatures T$_{\rm bb}\simeq$100-200 eV
 and luminosity 10$^{36}$ erg s$^{-1}$ at M31 distance do exist.
 I discuss the possibilities that most QSS may be SNR  in M31, or
 foreground neutron stars.  Two X-ray sources with both a soft and 
hard component are in the positions of a recurrent nova
 and another object that was tentatively classified
 as a symbiotic nova. These two sources may be black hole transients. 
  
\end{abstract}


\section{Introduction}

One of the main discoveries of  Einstein and especially of
  ROSAT were the {\it supersoft X-ray sources}
(hereafter, SSS).  SSS emit
detectable X-rays only at energy below 1 keV, with very large luminosity
  in the range 10$^{36}$-10$^{38}$ erg s$^{-1}$,
and their spectrum is approximately fit with a blackbody
at temperature in the range 150,000-1,000,000 K.
Since these sources are very luminous,
but very ``soft'', they are much more easily detected in Local Group
 galaxies,  towards which the column of neutral hydrogen N(H)
 is low, than in the Galaxy (see Greiner 2000 and references
 therein). 
The most luminous SSS are even detected outside the Local Group
(Swartz et al. 2002, Di Stefano \& Kong
2003,  2004, Kong \& Di Stefano 2003 and 2005).
 Even if the whole sample of SSS,
selected on purely phenomenological criteria, is not a
homogeneous class,
 there is evidence that a large fraction of these sources in the Local
Group   are extremely hot, accreting white dwarfs (WD) in close binaries, 
 burning hydrogen in a shell (see Kahabka \& van den Heuvel 
1997).  Close binary SSS (CBSS) include post-outburst recurrent
 and classical novae, the hottest symbiotic stars,
 and other low mass X-ray binaries 
 containing a WD, with typical orbital periods between
 4 hours and 1 day.  The recently rekindled debate on the
nature of the progenitors of type Ia supernovae (SNe Ia) focuses often on
this class of sources, as prototypes of single degenerate
 binary SNe Ia progenitors  (e.g. van den Heuvel et al 1991.,
King et al. 2003, Yoon \& Langer 2003, Starrfield et al. 2004). 
 SSS appear often to be transient or recurrent in X-rays, or 
 at least variable in X-ray flux (e.g. Greiner et al. 2000, 2004a). 

Most CBSS are indeed expected to be variable.
There are basically three mechanisms for variability
of the binary SSS described above. 1) A 
periodic radius expansion is accompanied by variation in
 the mass transfer rate $\dot m$, 
and feedback in the nuclear burning rate (e.g. Cal 83, Greiner
 \& Di Stefano 2000),  recurrently causing increase the optical
 and UV luminosity to increase and the X-ray luminosity to decrease.
 2) Repeated thermonuclear hydrogen shell flashes
(unlike novae, without mass ejection) are expected
 in a regime of mass accretion rate, $\dot m$, that varies depending
 on the model, but is around 10$^{-8}-10^{-7}$ M$_\odot$ year$^{-1}$.
Since the upper limits on the X-ray flux before the new generation
of X-ray satellites were not very high and the observations of
 SSS were seldom repeated, 
 we know very little about thermonuclear flashes.
 3) At higher   $\dot m$, all the energy produced in thermonuclear burning
 is immediately radiated (Fujimoto 1982, Kovetz \& Prialnik 1994),
 and generally we expect to find persistent X-ray sources, which
 are the most likely candidates for progenitors of neutron
 stars born by accretion induced collapse (AIC)
and/or of type Ia supernovae
 (see Yungelson et al. 1996, Starrfield et al. 2004). However,
 at high $\dot m$ there is another mechanism 
 for variability. A ROSAT source, RX J0513.9-6951,
 is optically bright
and ``off'' in X-rays for the 140 days of a recurrent cycle, 
 then it undergoes a rapid transition in less than 4 days to
 the supersoft X-ray source stage, with decreased optical
luminosity. This stage lasts for about a month, until 
 in less than 2 days, the source returns to the previous
state (Reinsch et al. 2000). The optical variation is
only 0.8 mag in V, while the ROSAT PSPC count rate was
measured to vary by a factor of more than 20. According to Hachisu
\& Kato (2003) RX J0513.9-6951 burns hydrogen in shell at the
constant rate necessary to reach the SN Ia
explosion. When more mass transfer rate is triggered than
 the rate at which it can all be burned, the outer layers
 of the WD expand and mass accreted from the secondary
is suddenly lost in a wind, causing the X-ray off state and
optical brightening.  Independently of the final fate
of a particular system, observing and studying shell-hydrogen
burning   WD,
we obtain a glimpse on the evolution of  potential  SN Ia progenitors
 in general, a central problem
 of modern astrophysics. Even what we
 know about the acceleration of cosmic expansion depends on type Ia SN. 
 
 M31 is the most luminous and massive galaxy in
the Local Group, and as such provides a large
 selection of nearby extragalactic X-ray sources,
 including SSS. At visual wavelength it is almost five 
times as luminous as the other spiral M33, 
and almost 16 and 80 times more luminous as our two
 nearby satellites, respectively the LMC and the SMC
 (Sparke \& Gallagher 2000).    
The content of neutral hydrogen in M31 is estimated
 to be $\simeq 5.7 \times 10^9$ M$_\odot$, about
30\% more than the Galaxy. The central bulge is affected
by a relatively low column density
of neutral hydrogen, N(H)$\approx 8 \times 10^{20}$ cm$^{-2}$,
 and the surrounding region has only  
 N(H)$\simeq 10^{21}$ cm$^{-2}$,
as it is shown  Supper et al.'s (1997) simplified
 model derived from Unwin's (1981) maps. Most of the neutral hydrogen
 is concentrated in a thick star-forming ``ring
of fire'' around it, with N(H)=7.7 $\times 10^{21}$ cm$^{-2}$,
 but the gas extends to a large radius, and it has  a rather
 patchy structure, with both ``holes''
of low column density  and dark clouds causing higher column density
in small regions (see Davies et al., 1976, and Hodge 1981). 

Due to its large stellar population and many 
 regions with low N(H),  M31    is ideal for investigating SSS. The 
 M31 population was studied with  ROSAT by
 Supper et al (1997), Kahabka (1999) and Greiner et al. (2004),
 with  Chandra by De Stefano et al. (2004) and Greiner
 et al. (2004). XMM-Newton is the best suited
 satellite for this type of research outside
 the central region of the bulge (where the
 excellent spatial resolution of Chandra is very
important), due to the high effective area, high sensitivity and quite
 good calibration of the EPIC detectors in the
 soft range. 
 A comprehensive study of the XMM-Newton observations of SSS 
 in M31 has not been published. De Stefano et al.
 (2004) and Greiner et al. (2004) examined most 
 of the XMM-Newton exposures of M31, and recovered 
 detections of the Chandra SSS as well as some originally
found with ROSAT. Using criteria
 that are as equivalent as possible to  the ones adopted by these authors, I 
performed an unbiased search for  SSS in M31 in the 
XMM-Newton observations, using also two  
recent public images that were not yet available to the above authors. 
Moreover, I searched for common detections of  
a second large subset of ROSAT SSS identified by
Kahabka (1999). 
In addition to a variability study, one of my aims is
to make use of the large effective area and
 higher count rates of XMM-Newton to obtain as many broad
 band spectra as possible, to derive conclusions on the physical
 nature of these sources.
 
Last but not least, I also examined another class of sources,
 called Quasi-Soft-Sources (QSS) by Di Stefano et al. (2004), 
 found in M31 and in other galaxies (Di Stefano \& Kong 2004). These sources 
 were discovered relaxing
the search criteria of SSS based on the hardness ratios and extending
 them to sources with slightly harder spectra. Di Stefano et al.
(2004)  define QSS spectra as 
  blackbodies with temperatures in the range 100-350 eV. 
It is important to bear in mind that this class of sources are not
 necessarily related to SSS, nor is there any evidence yet that some
of them may  be accreting WD.  Using the XMM-Newton data,  
 I also investigated the nature of QSS: are they really a new
 kind of X-ray sources with uniform physical characteristics ?    

\section{The data set: the XMM-Newton observations of M31} 

A description of the XMM-Newton mission is found in
Jansen et al. (2001). The satellite
 carries three X-ray telescopes with 5 detectors, which
were all used: EPIC-pn (Str\"uder et al. 2001,
two EPIC-MOS (Turner et al. 2001), and two RGS gratings
(den Herder et al. 2001). Due to the faintness both
in X-rays and in optical of the sources  studied, 
I could not use RGS data. Images were taken with the optical
telescope with the Optical Monitor (OM) 
 during a large part of the time, but they 
 are either not deep enough to detect the counterpart of the large majority 
 of the sources in M31, or they are effected by source confusion.
 The data were reduced using the ESA
XMM Science Analysis System (SAS) software, version 6.1.0,
 and the latest calibration files available at the time
of the analysis. The spectra were analysed with the
XSPEC software package (Arnaud 1996) not only with
 the standard XSPEC models, but also 
 WD atmospheric models  studied for novae in the LMC
 (see Orio et al. 2003) and in the Galaxy (Rauch et al. 2005).

The beautiful XMM-Newton images, plotted on the
contour  of the Galaxy, can be 
 seen in the survey article by Pietsch et al. (2005a). 
The parameters and characteristics of the exposures
are summarized in Table 1. The position of 
of the image centers are given by Pietsch et al.
(2005a). Four pointings of the central region of the galaxy,
 in full frame mode with a radius
 of $\approx$ 15 arcsec, were  repeated every 6 months for 2 years,
 with integration times
ranging from 3 to 18 hours. These exposures do not completely
overlap.  Also five regions with  center around the major axis of the disk
 were observed, three
 in the North and two in the South of the galaxy, 
 for about 18 hours each.  
Finally, one region was observed
in  the M31 halo for about 3 hours. All together,
the area was 1.24 square degrees, 
which is small compared to 6.3
 square degrees covered with ROSAT, but larger
than the 0.9   square degree observed up to now with Chandra. 
Table 1 reports the dates of the
observations, the nominal length of the exposure,
the amount of effective exposure used after the intervals
of background flares were excluded, the number of SSS observed in each
 exposure, the faction of detected previously known SSS over
 the total number observed, and the same ratio for QSS.
 The ``EPIC chains'' were run again with the updated calibration
 without using using special screening criteria (no
``low threshold'' at lower energy than 0.2 keV was set);
 this choice may yield systematically lower count rates than 
 those measured by other authors using a different screening
procedure.   
Since I was interested only on the softest sources,
 in order to exclude background and improve S/N I used
 only ``single'' events (PATTERN=0), and the conservative screening
criterion FLAG=0. 
  
During the time it took to complete this
 work, a catalog of M31 XMM-Newton sources
 has appeared (Pietsch et al. 2005a).
For the faintest sources, the count rates I measured  may differ from the
catalog entries by 2-3 $\sigma$. In the catalog
 the background was chosen in a semi-automatic way, but
there are often near-by sources
 and background fluctuations, which introduce complications.
Moreover, many sources
are faint and the count rates are measured with large error bars.
Focusing on the problem of supersoft sources for which
the background choice is more critical, I tried
to pay particular attention to the background extraction for every single source.
 The authors of the catalog ``flagged''
 18 SSS chosen with hardness ratio criteria that are suitable for XMM-Newton
   EPIC-pn but not for Chandra, and imply that virtually all photons
   are detected at energy below 0.5 keV. 
In another very recent paper, in observations done with XMM-Newton,
 Chandra and ROSAT, 21 X-ray sources observed with
these satellites are identified with
recent novae in M31 (Pietsch et al. 2005b). The authors
 of this article include even 2 $\sigma$ detections
 and have accomplished detections of the faintest
sources with a method to combine the information given
by the pn and  and by the MOS together, on order
to detect extremely faint sources (Pietsch 2005,
 private communication). I also
 identified X-ray sources with some recent novae,  using more
``conservative'' criteria, and compare my conclusions
 with the ones of the above authors.

\section{Selection based on hardness ratio}

Since SSS and QSS have been recently been studied
in M31 with Chandra (Di Stefano et al. 2005),
in order to compare the populations of SSS and QSS detected
 with Chandra and XMM-Newton, a first step is to translate
the criteria that were used for Chandra into XMM-appropriate ones. 
It is important to bear in mind that the hardness
ratio criteria used to select SSS with ROSAT (e.g. Supper
 et al. 1997, 2001, and Kahabka 1999) were quite 
 different from the ones used with Chandra, due to the
very different energy range and spectral response of the two
 X-ray telescopes and their detectors. Since in the recent past and
in the future both Chandra and XMM-Newton are the instruments used
in studying populations of X-ray sources, using selection criteria 
 as close as possible to the ones adopted with Chandra 
ACIS-S is a priority.
 
I examined first the XRT+EPIC-pn hardness ratios,
since  most sources are faint and were detected 
only with this detector. The two MOS yield better 
 spatial resolution, and are very useful in observations near
 the bulge, but
  the lower limit on the luminosity of the detected sources is
reduced by a factor of 5-6. XMM-Newton  offers another great advantage
 for this study: a better calibrated response than ACIS-S
for the low energy range, 0.2-0.5 keV,
 at which most of the flux is detected for these sources. The 
absolute calibration of the EPIC instruments
is accurate to 10\%, while relative calibration
 of EPIC-pn and EPIC-MOS is thought to be
accurate to 1\% in all the 0.3-1.0 keV range. In the 0.2-0.3 keV
 range the background is very high, but does not effect
 significantly the soft end of the spectrum of an 
 extremely soft and luminous source like many of the SSS (see also
Orio et al. 2003).

Di Stefano et al. (2004) examined three spectral bands: 
S (0.1-1.1 keV), M (1.1-2 keV), and H (2.0-7.0 keV). 
 The first, basic criteria established
 by Di Stefano et al. (2004) to select a SSS is 
HR1=${ {\rm (M-S)} \over {\rm (S+M)}} <$-0.8, which implies S/M$>$9,
 and HR2=${ {\rm (H-S)} \over {\rm (S+H)}} <$-0.8, or S/H$>$9. 
Because EPIC-pn is much more sensitive than ACIS-S in
the S band, but not so
in the M band, and because of the high background (a ``price'' to
pay using a telescope with larger effective area), it
turns out that strict
``Chandra-equivalent'' criteria can be applied to SSS in the Magellanic Clouds
and in the Galaxy only  to the few brightest X-ray SSS in M31.

A Chandra ACIS-S ratio S/M$>$9 translates in S/M$>$30
 and S/H$>$19 for all possible
models relevant to SSS, even considering a possible low 
luminosity ``hard tail'' due to
thermal bremsstrahlung or even  non thermal emission mechanisms. 
Let us suppose for instance 
 a blackbody-like source at T$_{\rm bb}$=50 eV and column density 
 by N(H)=$10^{21}$ cm$^{-2}$ 
with an additional thermal plasma emitting at kT=0.9 keV and at KT=5 keV. 
If we measure S/M$>$9 and S/H$>$9 with Chandra ACIS-S, using PIMMS
 we find the equivalent EPIC-pn (medium filter)
harness ratios: S/M=$>$33 and S/H$>$20.
 The high background is a significant problem because 
 we have to be able to measure very low values of 
count rates M and H in order to ensure the perfectly Chandra-equivalent
 algorithm can be applied (especially
for the pn detector). Since typical 2$\sigma$ fluctuations on
the background level in the M and H band are around 4$\times 10^{-4}$ 
cts s$^{-1}$, we need  count rates S$>$0.013 cts s$^{-1}$
to make sure that the M and H values are statistically significant
measurements, but few sources in M31 have such high count rates.   

Note that   Di Stefano et al. (2004, see also Di Stefano \& Kong 2003) further 
classify sources into ``SSS-3 $\sigma$'' (if the combined uncertainties
on S, M and H make the hardness ratios also uncertain) and ``SSS-HR''    
(if (S+$\Delta$S)$>$9(M+$\Delta$M) and (S+$\Delta$S)$>$9(H+$\Delta$H)),
 which pose the same background-related problems. It is obvious that application
of the Chandra-appropriate algorithm for selection of SSS is
 background limited in the M and H bands.
 For the EPIC-MOS, the ``Chandra-like'' criteria
are a little less difficult to apply, because we only need
 to verify that S/M$>$14, but the MOS are even less sensitive than ACIS-S,
 and definitely fewer sources are detected with very low count rates. 

 For this reason, after carefully analysing what are the typical
count rates necessary for detection in the XMM-Newton exposures
of M31 and fluctuation of the background, I chose a different
criterion for sources
with EPIC-pn count rates $>$0.013 cts s$^{-1}$,
still using the same S,M and H band defined for Chandra by the above
 authors.
If the M and H count rates are less than 2$\sigma$ above the background,
I define a SSS a source for which S/M$\geq$9.3 and S/H$\geq$9.
I verified that all the SSS defined as such with both Chandra
and ROSAT, if observed and detected, were automatically selected 
as SSS in this way. (Note that all the ROSAT SSS are also 
defined as Chandra SSS, but not vice versa). 
 
To select QSS, also the criteria given in Di Stefano \& Kong (2003b) 
are limited by statistics and by the background if strictly applied 
to XMM-Newton.
I therefore define a QSS a source
with the hardness ratio HR3=(S+M)/({\rm total counts})$>$0.98.
I verified that in this way, the QSS of Di Stefano et al. (2004)
 are
consistently selected, and so are a few new sources that are 
approximately fit with blackbodies at kT$<$350 eV,
 which is the required for QSS by Di Stefano et al. (2004).
 However, as discussed below, some QSS spectra are
fit even better using a steep power law. All these ``new'' sources,
 except one, had not been observed with ACIS-S at the time Di Stefano
 et al. (2004) wrote their article. 

 Table 2 and 3 show 
the coordinates, proposed source identification, count rates 
measured with  EPIC-pn and EPIC-MOS-2, and the parameters
of the best spectral fit with either a blackbody or power-law model.
For the sources in Table 2, the M and H count rates are always
 negligible, and that they were not reported. 
Even of many H count rates even in Table 3 are not significant, 
being less than 2 $\sigma$ above the background, M and H
 count rates, if different from 0, are reported in Table 3
to allow comparisons with SSS and among different QSS.
 Note that in Table 1 the S count rate is given in the 0.2-1.0 keV range rather 
up to 1.1. keV, because adding the 1.0-1.1 eV bin, due to
the spectral response, adds only noise and does not increase the signal
for any of the sources.
Additional spectral models, not
included in the Tables,  were tested to fit the data and are
discussed in the paper. The possible counterparts at optical
and other wavelengths were searched using the 
Digital Sky Survey, the SIMBAD and Vizier data bases,
including catalogs linked by Vizier,
the Local Group Survey (Massey et al. 2001) and the GALEX archival
images of M31. 
SSS that are selected using of HR1 and HR2 values, but  
almost certainly belong to the foreground have been
excluded from Table 2 and 3. This group includes only 5-6 objects
 whose position is within a 3-4" radius around the coordinates 
of a star with V$<$13.5. Only one object in Table 2
may be a foreground star, the source r3-122,
at B=18.7 and R=18.4.  It appears to be too blue to be an AGB
star in M31, but probably too luminous to be a different type of star in M31. 
Because of the soft X-ray spectrum,
if it is in the foreground it may be a AM Her star.

In Tables 2 and 3, even if the absolute luminosity is derived with the 
best fit and error bars are as large as 50\%, it is 
immediately clear that only luminous sources are detected. 
 At an estimated  distance of 770 kpc (Sparke \& Gallagher 2001),
and even with the longest useful exposure of 64 kseconds, 
the lowest detectable flux is of order few 10$^{-15}$
 erg s$^{-1}$ cm$^{-2}$, which implies a luminosity of a few 10$^{34}$
erg s$^{-1}$. Since in the low energy
 range photoelectric absorption plays an important role, 
 the minimum unabsorbed luminosity that is actually detected
is almost two orders of magnitude larger.
The immediate consequence of this selection effect due to
 the distance, is that we observe many transient sources at maximum,
but cannot follow them into quiescence. Most low mass X-ray binaries
and all classical novae (see
Orio et al. 2001 for a discussion on novae) 
have quiescent luminosities
that do not exceed a few 10$^{33}$ erg s$^{-1}$ and
 they are not detectable even with XMM-Newton at the distance
 of M31.

 Photoelectric absorption is another very important selection effect. 
 No SSS were observed in the South-2/3 and North-2/3 exposures,
 and no QSS were observed in the exposures North-2,
 only one QSS in South-2: these exposures, especially the North-2 and South-2
ones, cover in large part 
the region of the ``ring of fire'', where N(H) is high (see map by
 Supper et al. 1997). The lack of SSS detected with high N(H)
is very significant because it indicates approximate upper limits
on the temperature of SSS, which seems to be only rarely $\geq$50 eV.
Fig.1 shows the simulated spectrum of a blackbody with a temperature
70 eV, bolometric luminosity L(bol)=$10^{38}$ erg s$^{-1}$, detected
in a 60 kseconds exposure with N(H)=7 $\times 10^{2}$ cm$^{-2}$.  
This source, having the typical effective
temperature of a Galactic nova-SSS in the post-outburst year
(e.g. Rauch et al. 2006), would be qualified as a SSS 
with the criteria adopted in this paper, but it would not
be classified as SSS using the restrictive  criteria of Pietsch et al. (2005a). 
 The difference
between this spectrum and other harder ones peaked at about the same
value (see SNR in Fig. 2) is the abrupt cut  at E$>$1 keV, which
would give a clue that this is actually a much softer and
luminous source than it appears. The count rate would be 
approximately 0.01 counts s$^{-1}$, however with the same column density,
 a blackbody at 50 eV would yield only a
count rate 0.0007 counts s$^{-1}$, and at 40 eV, only 0.0002  counts s$^{-1}$,
so the source would be undetected against the the background.

Given the variable absorption in M31 and the ``softness''
 of the SSS and QSS, it is difficult to draw conclusions on the spatial
 distribution of these two populations in M31. 
 However, in an exposure of a region in the halo (or rather,
 extended thick disk) no SSS were observed, out of a total 26 
X-ray sources detected in that exposure. Therefore, a preliminary
 conclusion is   
 that SSS are much more frequent in the central disk. 

\section{Very different types of SSS }

 Altogether, 20 SSS were selected using the hardness criteria described
 above: 15 are included in Table 2, while 5 are in the
 positions of known supernova remnants
 (SNR) and are listed in Table 4.  SNR constitute 
 a  quarter of all SSS found using the hardness ratio
criterion, a fact
that may be very relevant for other external
galaxies in which SNR are yet to be identified.

 Note that in the Tables  the coordinates of these sources are mostly
 determined using the XMM-Newton images, but if they differ from the
 Chandra positions by $\simeq$2-3'', I have  listed the more precise position
 determined with Chandra.
The tables also show  a proposed identification,
 the count rates measured with EPIC-pn and or EPIC-MOS-2, and  the
 parameters of two spectral fits, obtained
with either a blackbody or a power-law, or adding both models,
  and finally the unabsorbed luminosity  at a distance of 770 kpc derived from
 the best fit. If the sources were observed in the gap between
 the CCD's (the bulge exposures were not overlapping, so
 each source was observed  more than once)
or were in a region with many bad pixels excluded by the screening criteria,
 I give only the count rate of the detector that 
observed the source. The sources that are close to the bulge
 cannot be well spatially resolved with EPIC-pn;
  this is the case of the bright source no. 7 in Table
 2 (r2-12), but most of these sources are resolved with EPIC-MOS, 
with the exception of  a central
 region of about 2 arcmin$^2$,
 where the sources cannot be spatially resolved with
 any of the EPIC instruments. The supernova remnant
  S And falls into this category, and so does the QSS r1-9
(source no. 8 in Table 3). In Table 2, I included count rates
 for the Chandra SSS that were observed again. 
 Only  Chandra source r2-66 was not detected again; however, 
 the upper limits are not significant.
 The Chandra SSS r2-54 and r3-115 are listed in Table 3 as QSS.
 The peculiar case of r3-115 is discussed further below. 

  Only two models are included  
 in Table 2, but for every source, if enough counts were detected, 
 I also fit other models. For each source I always used 
 also  the Raymond-Smith model, suitable for foreground cataclysmic variables (CV),  
but no spectra seem to be fit  adequately with this model.
 WD atmospheric models and composite models, done by two different
 groups and with a variety of abundances and log(g), were also used to fit the
 spectra, described in Hartmann \& Heise (1997), Orio et al.
(2003), and Rauch (2005).   
The best fit indicated in Table 2 is quite
 unique and is obtained with reduced $\chi^2 <$1.2 except in the very few cases 
 that are discussed in the text. When a fit was possible
 with both pn and MOS spectra, I have attempted
 to fit all spectra simultaneously. In one observation of 
  source no.14 (r3-8), there is a systematic discrepancy in the fit
of  MOS and pn data, discussed in Section 6.2. 
 I note that the pn spectra are usually fit with
 higher value of N(H) and higher effective temperature than the MOS
spectrum. 
 
  Grating observations of sources in the Galaxy and in the LMC have
 shown that there are basically two types of 
 CBSS: those in which the extremely hot and luminous WD 
atmosphere is observed 
(e.g. V4743 Sgr, Ness et al. 2003) and those in which a very
 soft emission line spectrum due to a wind is detected (e.g. Cal
83, Greiner et al. 2004 and Orio et al. 2004), but  
it is difficult to distinguish the two type of spectra
 with broad band CCD-type spectroscopy. The jury is
 still out as to whether the second type of binary SSS 
includes only WD or also other compact objects,
 but it is unlikely that the puzzle will be solved observing
 such distant sources as the ones in M31. The
 two classes, however seem to differ in luminosity,
not higher than a few 10$^{36}$
 erg s$^{-1}$ erg s$^{-1}$ for wind sources (see also Bearda
 et al. 2003, Motch et al. 2003, Ness et al. 2005),  but mostly
 around 10$^{38}$ erg s$^{-1}$ erg s$^{-1}$  for the WD
 atmosphere (e.g. Balman et al. 1998, Rauch et al. 2006, Lanz
et al. 2005).

\section{Supernova remnants}
 A difficult and important questions concerns 
 sources that are not close binaries, and mostly not even stellar, but
 cannot be distinguished from binary SSS from the hardness ratio alone. 
The   SNR or SNR candidates listed in Table 4 are have an extremely
 soft spectrum. The Table includes only those sources
for which  there is a (usually
published) optical, UV or radio identification
in the spatial error box (see reference column). Three of the sources have
 also been suggested to be SNR by Di Stefano et al. (2004) and others
 by Pietsch et al.  (2005).
  Five of the sources in the Table have hardness ratios
  that are typical of SSS according to the
 definition adopted in this work; the others are 
 QSS according to the definition adopted here.
 Table 4 includes also two SSS for which the XMM-Newton count rates
 are highly uncertain and are not reported: 
 S And, whose spectrum in the Chandra exposure
 is the softest of all, but it is not easily and ``cleanly'' extracted with the
PSF of XMM-Newton in the extremely crowded region of the bulge,
 and s2-42, detected in the XMM-Newton observation but partially obscured
by the junction between CCD's (see Di Stefano et al. 2004 for
 Chandra data). It is significant that the SSS-SNR in Table 4
objects bring the total
 number of SSS to 20 (in addition to the ones in Table 2): 5 SNR 
 out of 20 objects is a large number, corresponding
 to 25\%. This definitely demands further 
investigation not only of unidentified
M31 sources, but of other extragalactic SSS.  

Three SNR observed with ROSAT are included in Table 4, but they
 were  {\it not} selected as SSS in any ROSAT list, because the ROSAT-PSPC
 bandpass was different and
different hardness ratios were considered. Using 
the  S energy band from 0.1 to 1.1 keV (or up to 1 keV for our
XMM-Newton observations) 
to define the selection algorithm is a reasonable approach
using the Chandra ACIS-S detector,
but it makes it more difficult distinguish stellar, binary
SSS from the softest SNR. 
 Being able to use a lower energy limit for their ``soft'' band  
Supper et al. (1997, 2001) were able to exclude most SNR (but not all)
 from their sample. Examples
 of the soft SNR spectra are shown
in Fig. 2. Comparison with the other
 spectra shown in other figures in
this paper illustrates well that  the ``soft SNR'' spectrum 
  is usually not as soft as the spectra 
of the other sources shown in the other Figures
 in this paper (with some exceptions, like
 the very young remnant S And). In external galaxies, for which we 
need the spatial resolution of Chandra, yet cannot 
collect enough photons using a narrower range than the S range, 
the SSS or QSS samples probably include several unidentified SNR
(SNR data bases are not complete).

 At least two components are often necessary
 to fit the spectrum of the soft SNR. Since the ratio of soft to hard
photons measured with the EPIC detectors
is much higher than that measured with
 Chandra ACIS-S (and even much more than the one
 measured with Chandra ACIS-I, with which
 a few SNR were observed),   the softer component
appears much more dominant in the EPIC observations. 
 For this reason, with  low statistics the Chandra and XMM-Newton best-fits
  are often different. Only for source r3-63 the statistics
 are sufficiently good for a somewhat meaningful spectral fit.
The best fit to the  XMM-Newton spectrum is a Raymond Smith
+ power-law model (yielding reduced $\chi^2 \simeq$1, see Fig 2, upper panel).
 I also attempted fitting other models to this source: 
 blackbody, power-law, blackbody + power-law, Raymond Smith and
non equilibrium ionization (NEI, the last two models were found to be
 appropriate for the Chandra spectrum by Kong et al., 2002), but 
 a composite model best
fits EPIC-pn spectrum. The blackbody model alone (see 
parameters in Table 4) for r3-63 is  clearly
not suitable, yielding a best fit with reduced $\chi^2$=1.5.

However, very soft and faint SNR are generally fit well 
with one component, either
 a blackbody or a Raymond-Smith with the same luminosity.
For instance, the Chandra ACIS-S spectrum of r3-67 was 
fit by Williams et al. (2004) with a blackbody + power-law,
with  N(H)=5 $\times$ 10$^{21}$ cm$^{-2}$
 and T$_{\rm bb}$=50 eV.  The softest component is so dominant in the XMM-Newton EPIC-pn
observation, that a blackbody alone (even if at
higher temperature) is sufficient to fit the spectrum. 
The parameters obtained are
 N(H)=4.2 $\times$ 10$^{21}$ cm$^{-2}$  and T$_{\rm bb}$=130 eV
(indicating
intrinsic absorption, see right lower panel of Fig. 2). I also note that
the luminosity obtained with my best fit 
 is almost 8 times lower, although the value of Williams et al. (2004)
is still within the 3 $\sigma$ uncertainty of the
spectral fit, because of the low statistics.
 The spectra of r3-84 (proposed to be
the SNR Braun 95 by Kong et al., 2003) and of source no. 3, shown
in the left lower panel of Fig. 2,  are two other
 examples of SNR detected with low S/N. 
 Even if  their spectra seem to be more structured
than a simple blackbody, I cannot determine a more 
 appropriate model. The lower the count rate is, the broader
 are of course the 2 $\sigma$ contours in the parameter space.
For source no. 11, detected with only
 3.11$\pm$0.46 cts s$^{-1}$ in the pn image, the value of N(H)
varies between 1.4 and 8.5 $\times 10^{21}$ cm$^{-2}$,
 and the blackbody temperature between 73 and 186 eV,
 while the constant that determines the bolometric luminosity is
 unconstrained. 

Source no. 7  is included in the SNR table because it is a non variable
source, spatially coincident with both a ROSAT one and a radio source. 
 Even if  the count
 rate appears to vary within the 3 $\sigma$ level, this 
is surely due to the very off-axis position of the source 
 the core exposures. A Chandra position is missing
(this source was not observed), but the SNR
 identification has been proposed also
 by Supper (1997) and Pietsch et al. (2005). Two reported
 detections of different  planetary nebulae  in the catalogs
near this position,
 have been rejected and are now thought to be mis-classifications
of the extended SNR (Roth et al. 2004).

Table 4 illustrates well that many SNR detected in external galaxies with
low gas content often have, even if only in
 first and crude approximation, blackbody-like spectra and
using broad-band hardness
 ratios may be easily classified as SSS or QSS. However, with 
 Chandra at least   two SNR in M31 are 
 spatially resolved, and this is of course
 the best way to distinguish them from stellar point 
sources (Kong et al. 2002, 2003). Outside the local group, one 
 cannot even rely on spatial resolution. For SSS samples
in external galaxies, the starting point
to distinguish SNR 
 would be to obtain both  Chandra observations (to determine the coordinates and
 offer a possibility of identification at other wavelengths) and
 XMM-Newton observations of the same field (if source confusion
is not a problem in the X-ray images) in order 
to use a narrower range than ``S'' and  better assess how soft the source
really is.

\section{Are there background sources in the sample?}
Other spurious sources may be background AGN, especially 
if they are at low redshift, or/if they are narrow line Seyfert II galaxies,
  which often have a very soft X-ray spectrum (see Sulentic et al. 2000).
One of the ROSAT SSS was identified optically; preliminary results  on the 
optical spectrum indeed shows the typical signatures of an AGN at
 redshift z=0.187 (Greiner 2005, private communication).
   By fitting the spectrum with a few different models 
 and letting N(H) vary, in principle it is possible to 
 select possible candidate AGN among SSS and QSS, mainly because
 of high column density N(H) and apparent low luminosity at M31 distance
 (unless the source is in a dark cloud of M31),
 and possibly because of variable N(H) as well. However, 
 since both close binary SSS and AGN are often variable, time
 variability exclude SNR but not AGN. I have not found
 any source in Tables 2 and 3 that seems to show indications of being
an AGN, although this result is only preliminary,
 pending radio or optical identifications.

\section{Time variability of supersoft X-ray sources }

 In Table 2, 7 out of 15
 entries correspond to sources previously observed and classified as SSS
with Chandra (the positions are known with 1'' precision).
For these sources, for the ones detected with ROSAT and
 for the new ones repeatedly observed in the center, one can try to
investigate the variability on time scales of months
 and years. Among the Chandra sources, which are the most
interesting because of their precisely determined position,
 only one was not detected also with XMM-Newton,  
 source r2-66, but the upper limits on the flux
 are higher than the Chandra detection.  
Source r3-8 (see Section 6.2) is one of two non-SNR sources
 detected with XMM-Newton, that were listed as
ROSAT SSS in the additional compilation by Kahabka (1999)
 and were not examined by Greiner et al. (2004).
 For most sources observed with both Chandra ACIS-S and XMM-Newton,
a simple conversion
of the count rates with PIMMS is not sufficient to assess variability
due to the low count rates, large error bars and 
because of problems with the the ACIS-S calibration in the lowest energy
range. However, the four most luminous sources  are clearly variable.
 Their behaviour is discussed in Sections 6.1 and 6.2.
Among the entries in Table 2, 6 sources
 were not previously  detected before in X-rays, 2
 are very bright but transient,  and 5 are novae.

Greiner et al. (2004) found a high rate of failed
 repeated detections of SSS in the Chandra and XMM-Newton images
of M31, and inferred variability on time scales
of months. These authors examined  several SSS in the original
 classification by Supper et al. (1997) and
 they did not not include objects from the the additional list of  Kahabka
(1999). However, 
in the XMM-Newton exposures of the M31 central bulge
only 33\%  of {\it all} the sources previously observed
with ROSAT, Chandra  and/or XMM-Newton
(including also all those listed by Kahabka et al. 1999) were not detected
again. 8 out of 12 previously known SSS are in fact still bright in the longest
 bulge exposure done with the thin filter ``Core-4''. (Note that one
 of the 8 sources, however, is detected only in this exposure as a relatively
 ``hard'' source, and I discuss below even the possibility
 that this may be another, casually spatially super-imposed
 source).  To put things in perspective,
 we must also remember that the  flux detection limits
 in the second ROSAT survey were about 10 times lower than 
in the first survey.  This facts explains several
of the failed repeated detections
with ROSAT itself (see discussion by Greiner et al. 2004).
 Most important, the ROSAT  SSS examined by Greiner (2004),
were not close to the center of M31
(where source confusion was preventing ROSAT detections).

 Consistently with Greiner et al.'s (2004) result, 
the rate of failed repeated detections with XMM-Newton
 increases to $\simeq$70\% away from the central region: only 4 out of a total 
13 previously known ROSAT  SSS in the ``North'' and
 ``South'' fields re-observed away from the central region  
with XMM-Newton were detected,
 even if the exposures were quite long. Is this due to 
 intrinsic variability of the sources? Probably not in most cases, since
the higher rate of missed repeated detections 
in the regions that are far from the bulge 
 is explained by at Supper's (1997) schematic
 N(H) map based on Unwin's (1981)
 maps. These peripheral regions have much higher gas content than the
one within one  arcminute from the central nucleus. Even
 a very small variation in effective temperature and luminosity 
 may prevent detection with such a high column density.
It also appears from Table 2 that all the count rates
of the faint Chandra sources fluctuate  by up to $\simeq$30\% ,
 however the variations remain within the 3 $\sigma$ level except for
source r2-56 and, unusually, three novae, which seem to be genuinely variable
(not including the nova of 2000, which was not in outburst in the first
two exposures of the bulge). 
Whether the the fluctuations in count rate
are real, due to small  amplitude variability with a time scale of months,
or an effect of statistics, they are sufficient to 
make the  faintest  SSS undetectable,
 if they were re-observed at a larger off-axis angle or for less time.
A large fraction of SSS are intrinsically faint and only
 marginally detected at M31 distance, so they often
fall below the detection threshold when they are
observed again. In fact, almost all the sources 
 that were not detected again have low count rates,
 except the genuine ``transients'' discussed in the
 next Section.  

\subsection{Two transients: variable on short
and on long time scales}

 There were two  SSS in outburst during the 2 years of XMM-Newton observations,
 one was
reported at the time of the observations by Osborne et al. (2001) and
 the other by Shirey  (2001).  Both 
transients were observed only once; upper
 limits on the count rates are 50 and 10 times, respectively,
 lower  for other observations. In both case the best fit
luminosity is L$_{\rm bol} \simeq 3 \times 10^{38}$ erg s$^{-1}$,
 at the Eddington limit. 
 
The first transient, source no. 11 in Table 1, was described by Osborne et al.
(2001). It was  very luminous in June of 2000, yet
it had fallen below the threshold detection
of EPIC pn 6 months later,
in December of 2000 (only the possibility of a 2$\sigma$ detection
is suggested by Osborne et al., 2001). After a year
the upper limit to the count rate is lower than the 2000 measurement
by about two orders of magnitude. This source is only $\approx$10'' 
away from another one and cannot  
be studied in the EPIC-pn image, while in the EPIC-MOS images,
the sources 
 are separated fairly well.  A modulation with
a period 865.5 s (14.4 minutes) is clearly detected in the data.
King et al. (2002) discussed the possible models to explain
 this modulation, including a short orbital period double degenerate system.
These authors came to the conclusion that this is the spin period
 of  a magnetized WD in an intermediate polar system,  
and that hydrogen is burning steadily in a shell on the WD.    
Because of the periodicities that were later found for V1494 Aql
(Drake et al. 2002) and V4743 Sgr (Ness et al. 2003, Leibowitz et al.
2005), I note that the period of this M31 source
falls in the range commonly observed for post-nova SSS 
(Drake et al. 2002, Ness et al. 2003), or for the PG1059 WD 
that emit supersoft X-rays (e.g. Vaucler et al. 1993), attributed to
non-radial g-mode pulsations of the WD. 

 The other transient, source no. 12 is one of the softest
(see Fig.3). It was observed in June of 2001
and it was already below the detection limit of a 38 ksec
observation done on October 5 2001 with Chandra ACIS-s, and it was
not detected in January of 2004 with XMM-Newton.
This source seems variable during
the observation ($\simeq$90\% probability of variability),
 but I  could not detect   any periodicity (this is discussed
more extensively in a forthcoming paper on the
time variability of  SSS, Orio et al. 2005). 
 
 The spectra shown in Fig. 3 are  not very well 
fit with a black-body model.  The best fit is obtained with $\chi^2$=1.40
 and $\chi^2$=1.36 for source no. 11 and 12 respectively (see Fig.2).
With WD atmospheric models, however, the value
does not improve for source no. 12, while $\chi^2$=1.10 is obtained
for source no. 11, although with a very low value 
 N(H)=2.4 $\times 10^{20}$ cm$^{-2}$, which would imply 
that it is  a nearby Galactic
 source (but a possible Galactic counterpart should be
 detectable if this was true). 

If these two transients belong to the class of binary SSS, 
three possibilities may explain their time variability:
 optical novae in M31, a limit cycle like RX J0513.9-6951
 (Reinsch et al. 2000), 
or thermonuclear flashes without mass ejection. Although
 both sources were observed in June, when M31 is not
observable from ground based telescopes, it takes 
usually 6-8 months after the optical maximum for classical novae to reach
the X-ray supersoft stage (e.g. Orio et al. 2001b, Ness et al. 2004).
  This would mean that the optical outburst occurred
 when M31 was observable optically, but it seems 
unlikely that the novae where not observed in optical surveys,
 especially Macho surveys,
being carried on at that time (e.g. Riffeser et al. 2003). 
These  transients may be  like RX J0513.9-6951,
undergoing the limit cycle studied by  Hachisu
\& Kato (2003).
However, the second transient became bright at UV wavelengths
in OM images (Shirey 2001) when it brightened also
in X-rays, suggesting a different behaviour than
RX J0513.9-6951. Measurements at optical
wavelengths would be the key to understand this cycle and are missing. 
A variation in X-ray flux by an order of magnitude
or more may be due to a thermonuclear flash, 
 at $\dot m  \geq 10^{-8}$  m$_\odot$ yr$^{-1}$
(see Fujimoto 1982, Starrfield et al. 2004). 
Even without mass ejection, a thermonuclear flash causes
 expansion of the WD envelope and an optically
bright phase follows the X-ray brightening.
This variation is expected to be several magnitudes, much more than
the small amplitude of  RX J0513.9-6951, although not quite as large
 as for a classical nova.  Only X-ray and optical 
 monitoring done for several years would prove that we have the
 right time scales to infer the flash, but it seems that the 
X-ray bright phase, i.e.
the time to reach flash conditions, for most reasonable system parameters 
is longer than three months (e.g. Fujimoto 1982,
Kahabka 1998). In any case, monitoring the 
region near the center of M31 semi-simultaneously
in optical and in X-rays is the key
to solving the puzzle, and it will be very important to assess
 whether these two sources are really accreting and burning
hydrogen at the high rate required by  SNe Ia models.      
 
\subsection{Two bright, variable sources: ejecting mass in  a wind?} 

 Two bright and variable  SSS in Table 2 deserve special
 attention. The first one is source r3-8, whose broad band spectra 
 are shown in Fig. 4. This source is likely
 to be similar to the two transients described above, because even if
 it was observed every single time, the count rate varies
 by more than an order of magnitude.
A bright UV source appears to be at 304 arcsec from the 
the Chandra position  
 in the archival public GALEX images of M31, both in near and far ultraviolet
(the position uncertainty is 6'').
Imaging the field with the 3.5m WIYN telescope I found that the
closest star to the Chandra position is a  very blue star at B$\simeq$22.7
and R$>$23.  An initial astrometric solution shows that
it is at only $\approx$5''  angular distance from the Chandra
position  (see Fig. 5), and there is no star with B$<$23.5 closer
 to the Chandra position. Improved
astrometry and photometry will be included with new data
in a forthcoming paper (Orio et al. 2005, in preparation).
The ratio of the distances to M31 and to the LMC translates in a magnitude
difference of about 6 magnitudes. This implies that at M31 distance, only
 two of the seven SSS in the LMC, Cal 83 and RX J0513.9-6952 would be 
observed to have B$<$23.5, so definitive conclusions are not possible
yet.
 
 This X-ray source appeared to be bright in June of 2000, 
then the luminosity decreased
 in the following year and it increased 
 again in January of 2002 (see Fig. 4). There is also evidence
 of periodic variability on short time
 scales. In the June 2000 observation,
the power spectrum has a peak at the frequency corresponding
to a 2761 s period, and an even higher peak at the frequency
 corresponding to its alias, 5523 s (this is
discussed more extensively in Orio et al. 2005, in preparation).
 If the latter is an orbital period, it is surprisingly short, 
1.53 hours. Of course there are other possibilities: spin
 of the WD, or non-radial pulsations of the WD.

 The broad band spectra, shown in Fig. 4 with the best blackbody fit,
(the 2 $\sigma$ range of the parameters  indicated 
in Table 5),  do not constrain the bolometric
luminosity. However, the fits  indicate that it is very unlikely that the higher flux
and harder spectrum of the ``high'' states
in June 2000 and January 2002 are produced by a source at higher
 temperature. The spectrum of the June 2001 observation is 
best fit with a much lower value of N(H) than the other spectra.
 While the source luminosity reached a few  
 10$^{38}$ ergs s$^{-1}$ in the ``high'' states at the
 beginning and at the end of the observations,
 N(H) was high ($\simeq$2.3 10$^{21}$ cm$^{-2}$
 as opposed to only 7 $\times$ 10$^{20}$ cm$^{-2}$)
 obtained from the best fit of the third
 observation), causing the spectrum to appear harder. Note that
 for the last observations the pn spectrum fit 
 differs from the MOS one, and  indicates even higher N(H) and
 super-Eddington luminosity, $\simeq 2 \times 10^{39}$ ergs$^{-1}$.
The MOS best fit is the one reported in Table 2.  

 Fits with composite models
 and with several WD atmospheric models were also attempted, with 
 different abundances (see references above), but they do not yield 
 an improvement over the blackbody fit and do not indicate 
 a significantly different value of N(H). The 2 $\sigma$ contours
of the bolometric constant are not ``closed'' towards lower
 values, so the bolometric luminosity is not
 well defined. However, if we examine the best fit
 we may conclude  that, as the source luminosity decreased,
the effective temperature at first
increased or remained constant, and only in the third
observation there is marginal evidence that
 it may have have been lower.  N(H), on the other
 hand, must have decreased as the
 XMM-Newton count rate decreased.  A Chandra HRC-I observation
 shows that the luminosity was already higher again
by October of 2001 (the source was in the
 gap between CCD's in the ACIS-s observation
of the same period).

 In all the XMM-Newton observations,
and particularly  the ones in the faint state, even if a
 blackbody is a good fit there seems to be some structure 
 that is suggestive of spectral lines (see Fig. 4).
 A likely explanation for this spectral behaviour, if we are dealing
 with a WD burning hydrogen in a shell, is that thermonuclear burning always
occurs at an almost constant rate,  because it is regulated 
 by a wind, as in RX J0513.9-6952 . When the burning rate increases
 because more material is being accreted,
 probably as a consequence of irradiation of the secondary, 
 the luminosity exceeds the Eddington value. Then a wind ensues 
 (causing the higher N(H)), and strips the source of extra material to burn,
 until the WD shrinks, then cools,
 and the energy generation rate goes back to a level at which
all the accreted material is immediately burned. We observe at this
 point the lower luminosity and lower absorption spectra
 of December 2000 and June 2001, shown  of the right side
 of Fig. 4. In this scenario, there is a limit cycle like for
 RX J0513.9-6952, but with a very important phenomenological
difference: the wind is never completely
optically thick to X-rays,  so the source 
at first becomes even X-ray brighter,
 instead of disappearing from the X-ray window. 
 As in RX J0513.9-6952, it is unlikely that the 
wind strips the WD of a significant portion of accreted 
 material like in classical novae, so this source may be considered 
 to be a potential SN Ia candidate. 
As mentioned in Section 6.1 for the two transients that went below the detection
 threshold limit, quasi-simultaneous optical
 and X-ray monitoring of M31 is needed for a few years to 
 obtain the necessary information to fully understand 
how the cycle works.

 Source r2-12, for which a light curve in the range 0.5-7 keV
 was sketched by Di Stefano et al. (2004) under the assumption of
 constant blackbody temperature,  varies much less (only
 a 50\% amplitude in count rate once). 
 Even if it is very soft, it cannot 
 be fit with a single blackbody or with a WD atmosphere. As found by 
 Di Stefano et al. (2004), an additional power-law component
 is necessary to fit the spectra of three observations. 
The 2 $\sigma$ ranges of the parameters are given in Table 6.
For the data of December 2000 and of January 2002 I obtain a 
reduced $\chi^2\simeq 1.5$ for the best fit, suggesting
that even more components may be needed. 
 Only for the observation of June 2001,  
  a value of $\chi^2$/d.o.f.=1.1 is obtained by combining 
to a blackbody with a temperature of 62 eV an additional
multi-temperature disk (``diskbb'' in XSPEC), instead
of the power-law component.
   All the spectra seem to be structured and possibly 
have   several components.
The best fit value of N(H) again is higher, and the luminosity as well,
in the second year of observations. For the luminosity, there is 
no overlap of the 2 $\sigma$ contours in the observations
of these two years (see Table 6). In the last two observations the
luminosity is likely to have been super-Eddington 
 at M31 distance, a few 10$^{39}$ erg s$^{-1}$.

 The minimum
 bolometric luminosity obtained at the 3 $\sigma$ confidence level
keeping fixed values of N(H) and T$_{\rm bb}$ is 9.5$\times 10^{38}$
erg s$^{-1}$. 
Because of the large luminosity and the complex 
spectrum, if this source is confirmed to belong to M31,
it resembles the Ultraluminous
 X-ray Sources  (ULX) detected outside the Local Group,
in M101, that appear to be SSS in the most luminous state
(Kong and Di Stefano 2005).
 Perhaps it constitutes the ``missing link'' with the ULX class
 in the Local Group.
   
\subsection{Classical and recurrent novae, and black-hole candidates}

One important clue to the nature of  SSS is the 
frequence of classical and recurrent novae among them. It is clear
 that many novae turn into  very luminous  SSS for some time
 after the outburst, as the ejecta gradually become optically
 thin and as long as the WD is still burning hydrogen in a shell.
 For V4743 Sgr, the abundances appear to be typical of CNO processed
 material, implying that the post-nova WD  {\it at least in some cases}
 still burns the previously accreted, and not completely ejected,
 envelope material (as opposed to nuclear burning ignited
 in the external layers of the WD itself). If this hydrogen burning
 in a shell continues for a long time (i.e.  at least 10 years)  
 the leftover envelope mass is not negligible, and there is a possibility 
 that the WD grows towards the Chandrasekhar mass, since all novae
 are recurrent on long time scales. However, no nova has been observed
 to remain a  SSS for more than $\simeq$10 years (see Shanley et al.
 1995, Orio et al. 2003), and novae that burn hydrogen in a shell
 for several years are rare (see Orio et al. 2000).
 In addition to this, the objects that
could mostly contribute to the SNe Ia rate
are novae with short recurrence times, less  
than a century, i.e. Recurrent Novae, but they have never been observed
to undergo thermonuclear burning for longer than a month after
 the outburst (see Orio et al. 2005). 

 A nova in M31 was found to be a ROSAT SSS in a peripheral region of M31
 (Nedialkov et al. 2001).
 I have therefore overplotted on the XMM-Newton images of M31 
the position of 
 more than 200 novae, identified with H$\alpha$ surveys
of M31  in the last 25  years before the observations. 
 About half of these novae  were included in the XMM-Newton
 pointings. The list includes
 the IAU Circulars of the last 12 years, Astronomer's Telegrams
 of the last  2 years, and novae with coordinates reported
in the following papers: Darnley et al. (2004), Joshi et al.
(2004), Shaefter \& Irby (2001), Sharov et al. (2003 and 1992),
Sharov (1992) and a series of papers by
Sharov \& Alknis (1998, 1997, 1996, 1995, 1994) 
Ciardullo et al. (1987), and only a few the latest of the
 novae in Rosino et al. (1989). Most positions are
 known within only 1''  uncertainty. I have found that the position 
 of 5   SSS is coincident within about 3" with the coordinates of  
the list of optically identified novae, two of which are
Chandra  SSS, with coordinates determined with 1" uncertainty
 in the position, that
 overlap within less than 2" uncertainty with the novae position. 
The explosions of these X-ray identified novae have been recorded
in 1992, 1995 (2 objects), 1996, and 2000.
Not surprisingly, these were all novae whose outbursts
were recorded since 1992, that is in the 10 years
before the X-ray observations. Three spectra,
 and the comparison with a bight Galactic nova,
 are shown in Fig. 6. The spectrum of the Galactic nova,
 V4743 Sgr, was also detected with the
RGS gratings. It turns out to be atmospheric in origin
 and it has many deep absorption
 features, which however are totally unresolved
 with EPIC-pn (Ness 2004, Orio
et al. 2005, preprint).  The spectra have high S/N, but are generally well fit
 with both a blackbody at high luminosity
 (L$_{\rm bol}\simeq 10^{38}$ ergs s $^{-1}$
or atmospheric models and have temperatures
 in the range 30-50 eV . The spectrum of V4743 Sgr
 appears to be different because of
 a different N(H), probably a few times 10$^{21}$ cm$^{-2}$,
which decreases the flux detected below 0.4 keV, however
 like the M31 novae spectra,  it is truncated after 0.5-0.6 keV.
 It is interesting
 to note that Novae 1992-01, 1995-05 and 1996-05 (from the list of Shaefter \&
 Irby, 2000) appear to have fluctuations in count rate 
 that exceed the 3 $\sigma$ level.  This makes it necessary
 to have multiple observations before a nova is
 deemed to have ``turned off' as supersoft X-ray source, since the 
  X-ray lightcurve may have several fluctuations before turn off
 of the  SSS.

The same novae  have been recently reported to be X-ray
 bright in a preprint by Pietsch et al.
 (2005),  although Pietsch et al. also find XMM-Newton detections
 of three more objects which they identify with recent novae. I find that
one of them (AGPV 1576)
 is detected at a 3 $\sigma$ level in a Chandra-HRC observation
 (without spectral resolution), but I cannot confirm the detection above
 this level with XMM-Newton-pn. In the position of nova 
1997-09 of  Shaefter \& Irby (2001) and 
 Nova V1067  I cannot confirm a detection above the 2 $\sigma$ level.
 These additional detections of Pietsch et al. (2005) were made using 
a sophisticated method in order to detect the faintest sources
making use of the combined information of pn and MOS,
 and even a 2 $\sigma$ level detections are reported in the paper 
 (Pietsch 2005, private communication). My approach is more conservative,
I consider only detections above the 3 $\sigma$ level in the pn
images. In any case, the number of nova-SSS is a large
percentage ($\simeq$25\%) of the total number of SSS, yet a 
small fraction, about 5\%, of all the novae in outburst in the last 25 years,
which becomes only close to 10\% adding the 3 additional
candidates of  Pietsch et al. (2005).
    
I considered the intriguing  possibility of  identifying
 X-ray novae containing
a black-hole, that have recurrent outbursts
on few years time scales accompanied by optical outbursts
of large amplitude, often with $\Delta$m=8 mag, without
significant mass ejection. Could these outbursts be selected
 in the H$\alpha$ surveys that identify classical and recurrent
novae? Even if the
 H$\alpha$ line in outburst is much less prominent
over the continuum than that in classical novae, the total
luminosity of these objects in outburst is
comparable to classical novae and especially
 to recurrent novae,  so unless  subtraction is made with 
an image in an off band (usually, Johnson R),  taken at the same time
as the narrow filter exposure, both novae and X-ray novae 
 appear virtually similar in an H$\alpha$ image.

 Since X-ray novae
are recurrent, there is a small chance that an object
hitherto observed only optically may be in outburst
again during the XMM-Newton observations. Quiescent X-ray
luminosities for these systems are in the range
 10$^{30}$-10$^{33}$ erg s$^{-1}$, well
below our detection limits, and typically X-ray novae return
to quiescence within a year. 
The source r3-115, already listed in Table 2,
 shows a very unusual and surprising behaviour.
Riffeser et al. (2001) and Pietsch et al.
(2005b) consider it to be a possible symbiotic nova. It appeared 
as a typical, very soft  SSS in observations done with ACIS-S in 2001, 
(see also Di Stefano et al. 2004) 98 days after the C-3
 XMM-Newton observation.  However, it was never detected before,
and even the detection of a  SSS in the C-3 observation
(Pietsch et al. 2005) cannot  be confirmed above the 3 $\sigma$ level 
in the pn image (note that one of the two  SSS transients of Section 6.1
is displaced by  $\simeq$10'', making the detection of a nearby,
 soft and very faint source almost impossible).
However, the spectrum in the C-4 exposure, taken
 107 days after the SSS detection with ACIS-S, is shown in Fig. 6a:
 it has a lower blackbody luminosity, but shows now an important hard
component, fit as a power law. There is a chance of casual
 overlap with  a hard X-ray transient that cannot
 be spatially resolved, but since the source is
 away from the bulge, this
is unlikely. A source whose spectrum is becoming harder and
 less luminous in time  suggests the behaviour of an X-ray nova,
 with an outburst recurring every few years, casually
observed after an outburst with XMM-Newton.  A similar transition from
supersoft to harder X-ray spectrum was also observed in two peculiar
 sources, the Ultra Luminous  one in M101 (followed 
by Mukai et al. 2003 and  Kong \& Di Stefano 2003) and 
the SSS of M300 (Kong \& Di Stefano 2005).
 Di Stefano et al. (2004) identify r3-115 with a very
 red star at V$\simeq$22 in Hubble archival images, and
propose the identification with a symbiotic. In any case,
 the optical counterpart is likely to host a giant, which would be 
 unusual,  albeit not impossible for a black hole X-ray nova
system. 

The most interesting case is the one of recurrent nova Rosino 140 
(see Ciardullo et al. 1987), which is found in the position of the hard X-ray
 source r3-13 ($\alpha$=00,43,13.4 and $\delta$=+41,18,13),
 observed with Chandra and XMM- Newton. This source is not mentioned
in Pietsch et al. (2005) but it is a faint and probably it is 
 variable star. In exposure core-3 the source appears brighter
than in the other bulge observations, but a comparison is made
difficult by the presence of another faint source 6-7'' away.
In the MOS, it is easy to avoid the nearby source and extract a clean
spectrum, with a
count rate 0.0014$\pm$0.00031 cts s$^{-1}$,
 that appears relatively hard (see Fig. 6.b)
The statistics are very poor for a detailed spectral fit, but
it is clear that r3-13 has both a soft portion and also a   much harder 
and luminous tail than a classical nova. Since
 it is classified as a recurrent nova, and 
there is some evidence of X-ray variability, the possibility 
of a black hole transient should certainly be considered.

\section{Quasi-Soft Sources: a real new class?}

 In order to analyse the nature of the QSS, it must be
 remembered that no single X-ray observations
 ever went so deep imaging {\it any} an area of the sky
 effected by low N(H), allowing for the first time 
 the discovery of new, soft and faint sources.
Whether these sources belong to a uniform class 
is an intriguing problem.
 
 Table 3 contains 18 sources, most of which were previously selected by Di Stefano et 
al (2004). A few more with  HR3$>$0.98
 in at least one observation, detected at least or above the 3 $\sigma$ level,
 are included.  Three of the Chandra QSS were not detected again,
 but the XMM-Newton upper limits are
not significant to assess variability . For the Chandra source no. 8, r1-9, which
is in the inner nucleus of M31, using the XMM detectors  
 I can only assess the detection and give an upper limit to
the count rate, because source confusion cannot be
avoided. 5 more QSS are in table 4 because of their association
with a SNR: they represent 5 out of 23 sources.
  
  A close look at table 3 reveals that relatively
 high count rates were measured only for two sources,
r1-9 and r3-111. With the exception of source 
 no. 18, which as mentioned above may be Galactic, all the other non-SNR QSS have no 
detectable counterpart at other wavelengths.
For the sources detected with better statistics, a
 power-law with slope
 $\nu \simeq$2-2.5 fits the XMM-Newton EPIC-pn spectra better than
 a black-body model. The special case r3-115 has been discussed
in Section 6.3. I note that source no.4,
 no. 5 (r3-111), no. 9 (r2-62), no. 10 (r2-54) show some evidence
 of variability in count rate, although it exceeds the
 3 $\sigma$ level  only for for r2-54. For 8 sources 
a  blackbody at T$_{\rm BB}$=110-190 eV appears as a better
spectral fit  than a power-law, but they are all faint,
with maximum count rate (8.93$\pm 0.84)\times 10^{-3}$ cts s$^{-1}$ 
 measured with EPIC-pn. It cannot be ruled out that the blackbody
 fit appears to be the best because of the low
 statistics, like for the SNR in Table 4 (see discussion above).
 Comparing count rates and spectral fits in Tables 3 and 4, it seems that 
the faint, non variable QSS  may be SNR.
 I do not think that this is probably true for QSS in external galaxies,
 which seem to be intrinsically more luminous (Di Stefano \& Kong
 2004) but it is a possibility that cannot be ruled out for M31.
The database on SNR in M31 is probably not complete yet, and
this possibility would have important implications
for the supernova rate in M31. However, I note that a point {\it against} the
SNR hypothesis is however the lack of GALEX counterparts among
 QSS (only one  GALEX source among 18). Most known
SNR do have a UV counterpart at M31 distance. 

 If deeper observations reveal
 that the blackbody is indeed a good fit, the alternative possibility 
of Galactic neutron stars must also be considered. The lack
of optical counterpart, the blackbody temperature and
 the would-be luminosity at a distance $\simeq$8 kpc, 
$\approx 10^{32}$ erg s$^{-2}$, all fit this possibility.
 Finding several neutron stars at this high Galactic latitude
would imply a great re-heating efficiency, which is somewhat
surprising, but not impossible. Many cooling neutron
 stars are probably unidentified among the faint
 X-ray sources without optical counterparts,
 but comparisons with the known Galactic sample
 of such sources are not statistically
significant yet. Since the study of ``blank field sources''
 relies up to now mostly on ROSAT HRI data because a large
 field of view and a relatively precise position are
 needed, there is still very little spectral
 information on the faint neutron star
candidates (see Chieregato 2005, Chieregato et al. 2006).

  On the other hand,  if  part of the QSS are neutron stars
in M31, they  have large luminosities (a few times 10$^{36}$ erg s$^{-1}$),
 which would be consistent with anomalous X-ray pulsars
(AXP) rather than regular 
neutron stars (AXP however are usually less soft). 

\section{Conclusions}

 I have performed a systematic search of SSS and QSS in the XMM-Newton
  observations of M31. When possible, I have studied the spectra 
and the time variability characteristics  of these sources and searched 
counterparts at other wavelengths. These are
 a few conclusions that can be drawn from this study:

1) This ``close look'' M31 SSS population clearly
confirms that, as noted in other studies, using only hardness ratio criteria
 we  cannot select a uniform type of sources
undergoing the same physical mechanisms.

2) Using hardness ratios based on broad spectral
 bands like in Di Stefano et al. (2004), at least 11 SNR are selected
 as  SSS. However, SNR tend to have on average harder spectra than H-burning
 WD in CBSS, so the number of SNR among SSS is reduced if hardness ratios
 with narrower band-passes are used, like for ROSAT (Supper et al. 1997) or
 in the XMM-Newton catalog (Pietsch et al. 2005a).
  The presence of SNR in  the SSS
 data bases in external galaxies may be assessed not
only with multiwavelength observations, but especially complementing
Chandra observations, necessary for the good spatial resolution, with 
 XMM-Newton observations, in order to detect more signal in
the soft band. 

3) Post-outburst novae make up a  quarter of 
all  SSS, but only less than 1 in 10 nova of the last
 25 years is detected as  SSS. Up to now,
 in M31 like in the Galaxy and the in
the Magellanic Clouds, no nova was detected in X-rays after more than 10 years.

 4) 1 out of 10 novae in outburst in M31 is associated with a SSS,
indicating ongoing shell hydrogen burning.
 Yungelson et al. (1996) evaluated the SNe  Ia rate, the nova rate and the interacting
binary formation rate (e.g. Yungelson et al. 1996). Following
their reasoning, we  find that, if only 
 10\% classical and recurrent novae keeps on burning hydrogen in a
 shell for several years, these systems are not the main class of
progenitors of type Ia supernovae.
   
 5) Large flux variations by more than one order of magnitude,
consistent with the limit cycle of thermonuclear flash models
or with ``wind-regulated'' sources, are 
 detected in 20\% of those  SSS that are not SNR (3 out of 15 sources),
allowing for the possibility that they are CBSS, and that
 a relatively large number of CBSS
 may be burning hydrogen in a shell at the high rate required by type Ia supernova
 models,

6) At least one variable  SSS has a very large luminosity,
 of a few 10$^{39}$ erg s$^{-1}$ at M31 distance; 
the spectrum is surely not a simple blackbody and must be produced by 
 two or more components. This source, detected with Chandra as
 well, may be the missing link with ULX of external galaxies.
This is a very interesting finding. I dare suggest that monitoring this close-by
source in Andromeda we may be able to find a substantial clue to the nature of ULX.  
 
7) Looking at the present statistics, 
it seems that QSS may include a large number of SNR candidates,
 although it is puzzling that counterparts at FUV/NUV wavelengths are
not detected.  QSS may also include foreground neutron stars, or even
softer-than-usual AXP in M31.

8) Cross correlating the novae with with the positions of all X-ray sources
observed with XMM-Newton (not only X-ray sources), two non- supersoft X-ray
sources have been found.  Both are variable, one can be classified
 as ``transient'' and seems to be variable not only in flux but also
in spectral characteristics. These two 
objects are candidate black-hole transients.
 It will be interesting to monitor the X-ray behaviour
of  X-ray sources that are spatially
coincident with classical novae, not only 
in order to study H-burning WD, but also with the aim of 
selecting black-hole transients in the Local Group galaxies.
 
 I would like to conclude reminding that M31 gives us a
wonderful opportunity to obtain statistics of whole classes of X-ray
 sources, and that both XMM-Newton (because of the large effective
 area)  and Chandra (because of the low background and especially
 of the high spatial resolution, of unique value in the very crowded
central region) are exceptional instruments to study X-ray sources
populations and monitor the variable sources. 

\acknowledgments

I wish to thank Jay Gallagher for many useful  and
interesting conversations  and for
critically reading an early version of the manuscript, Emre Tepedelenlioglu
for his help comparing the XMM-Newton images with the Chandra ones, 
and Jochen Greiner, Hakk\i \ \"Ogelman and Wofgang Pietsch for useful
 discussions. 



%
\clearpage
\begin{figure}
\begin{center}
\includegraphics[width=11.5cm,angle=-90]{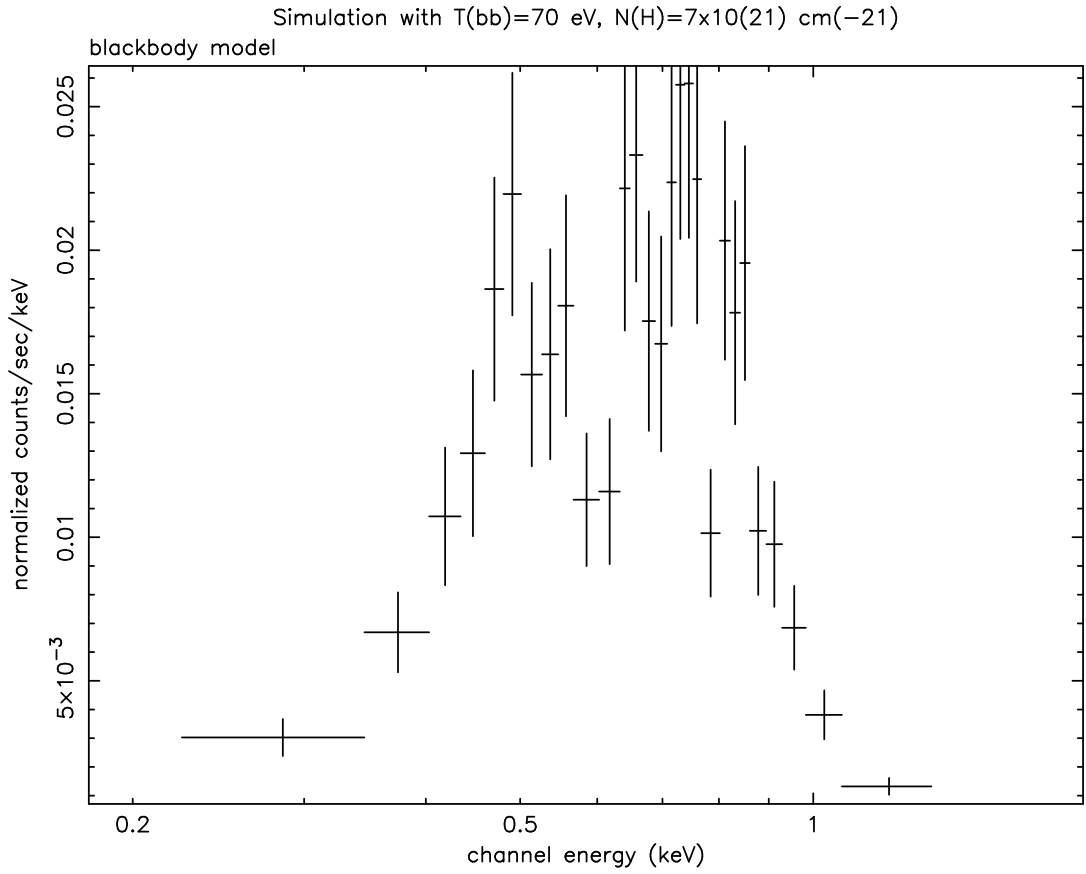}
\end{center}
\caption{\small {The simulated blackbody spectrum 
 of a source with T$_{\rm BB}$=70 eV, N(H)=7 $\times 10^{21}$ cm$^{-2}$,
in an 70,000 s exposure, comparable to ``South2'' and ``North2''. 
The spectrum appears to be still very soft, but very absorbed.
Note the abrupt cut at 1 keV.}} 
\end{figure} 
\begin{figure}
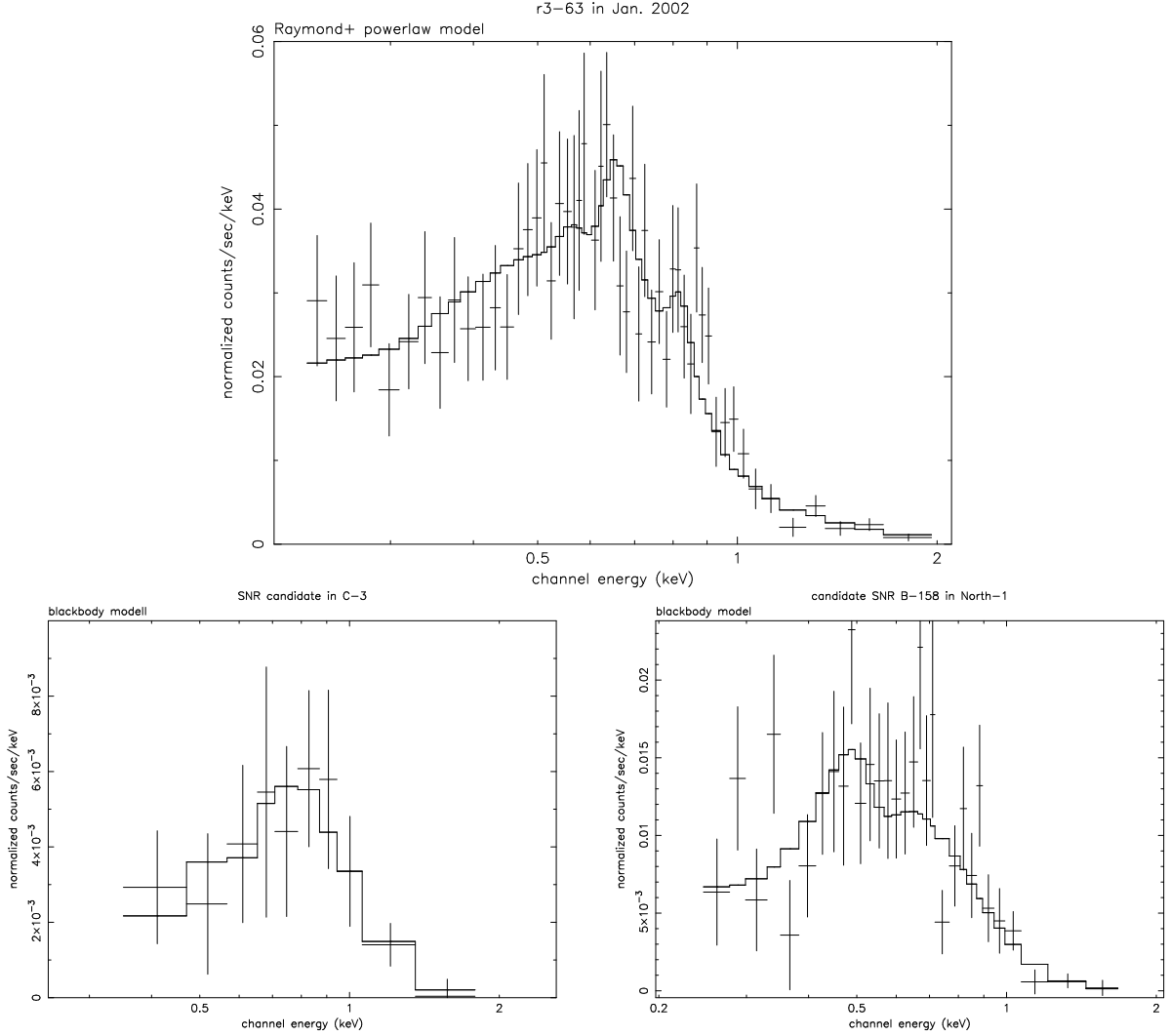

\begin{center}
\includegraphics[width=8.cm,angle=-90]{f2a.ps}
\vspace{0.5cm}
\includegraphics[width=6.cm,angle=-90]{f2b.ps}
\hspace{0.5cm}
\includegraphics[width=6.cm,angle=-90]{f2c.ps}
\end{center}
\caption{\small {The upper panel
 shows the spectrum of r3-63 (no. 5 in Table 4), the only supersoft SNR 
 with higher count rate, fit with a Raymond-Smith+powerlaw model, with
kT=244 eV, $\nu$=2.8, N(H)=8 $\times 10^{20}$ cm$^{-2}$. The unabsorbed
luminosity is found to be 1.4 $\times 10^{37}$ erg s$^{-1}$ in the
0.2-7 keV range. The two lower panels show 
the spectra of the two X-ray supersoft SNR candidates in Table 4, no.4
and no. 7, binned with S/N=20, and the best fit with  blackbody models
(with T$_{\rm BB} \simeq$120 eV)
are shown in the two lower panels.
 Even if the blackbody fit
is far from being perfect, $\chi^2=1$ and $\chi^2=1.1$  is obtained
 in each case respectively.
 Because of the low statistics,
 the blackbody fit appears to be better than 
more complex and perhaps physically appropriate models.}} 
\end{figure}
\begin{figure}
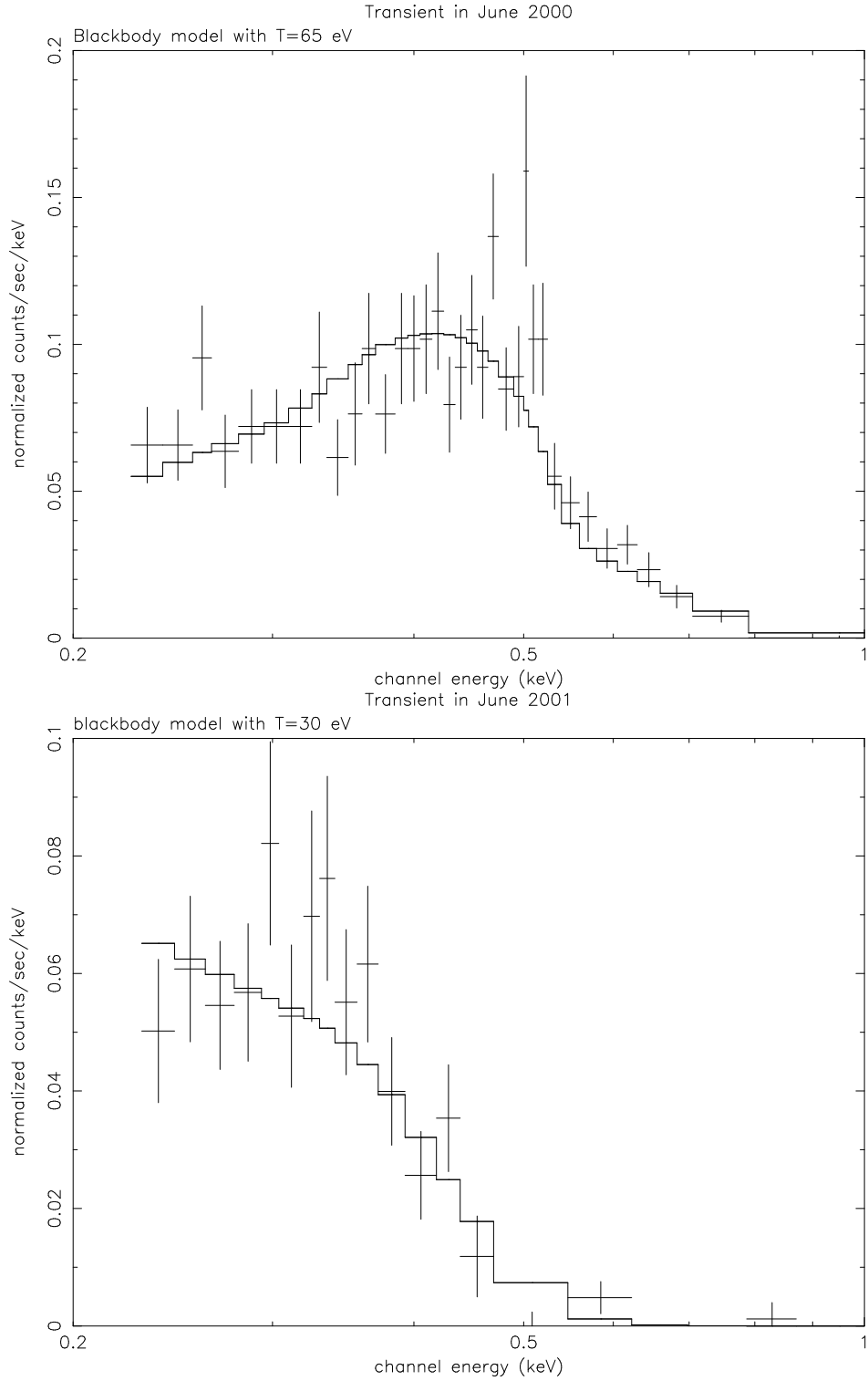

\begin{center}
\includegraphics[width=10cm,angle=-90]{f3a.ps}
\includegraphics[width=10cm,angle=-90]{f3b.ps}
\end{center}
\caption{\small {The spectra of the two transients, observed in June of 2000 and of 2001,
 respectively, and fit with a blackbody model described in Table 2 and in the text.
Note how the counts are truncated at some energy lower than 1 keV, 
which is a significant difference in comparison with the SNR spectra in Fig. 2.}}   
\end{figure}
\clearpage%
\begin{figure}
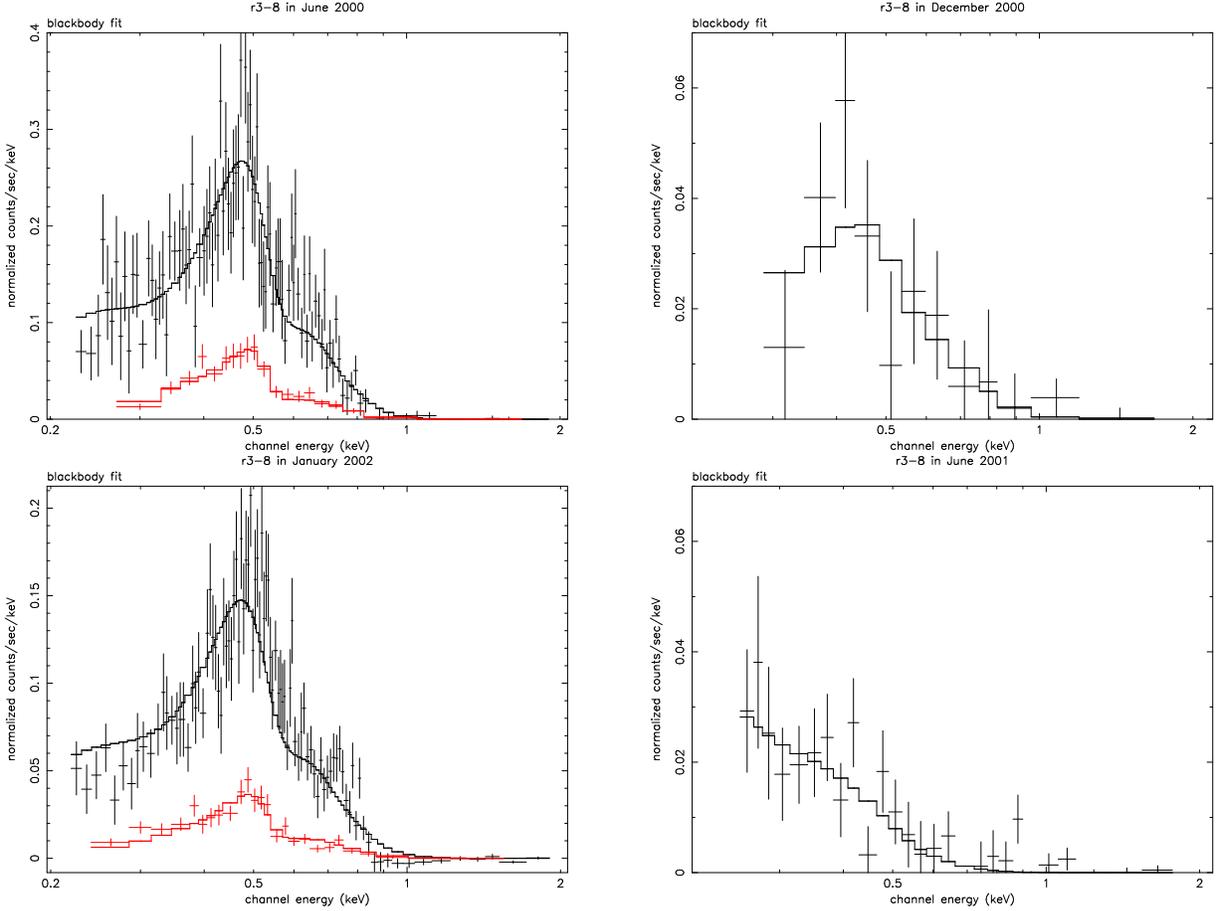

\begin{center}
\includegraphics[width=6cm,angle=-90]{f4a.ps} 
\hspace{0.8cm}
\includegraphics[width=6cm,angle=-90]{f4b.ps}
\vspace{1.5cm}
\includegraphics[width=6cm,angle=-90]{f4c.ps} 
\hspace{0.8cm}
\includegraphics[width=6cm,angle=-90]{f4d.ps}
\end{center}
\caption{On the left are the pn and MOS spectra of source r3-8 observed
 in June 2000 and January 2006, in ``high'' state, and best blackbody fit, 
on the right the spectra of the source taken during the ``low'' state, in
 December 2000 and June 2001. The best spectral
fits for the blackbody component indicate T${_{\rm BB}}$ 64 eV in the
first and last observation, 74 eV in the second and 48 eV in the last.
N(H) appears to have decreased in the second and third observation (see Table
2 and discussion in the text). }
\end{figure} 
\begin{figure}
\begin{center}
\includegraphics{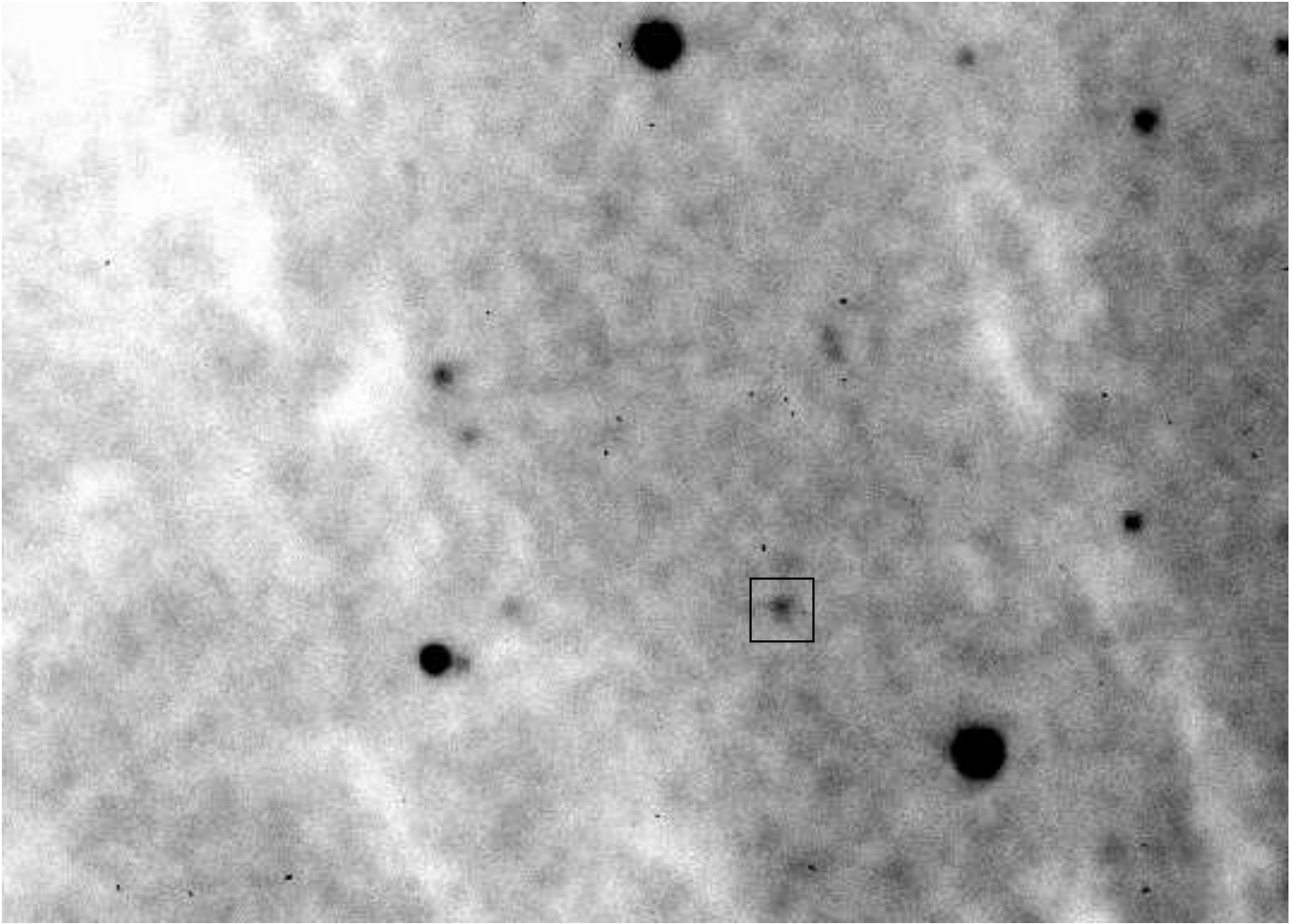}
\end{center}
\caption{An image of the $\simeq$1x1.3 arcmin field in which the
source r3-8 is included, obtained with 
the 3.5m WIYN telescope and the Mini-Mosaic imager on June 15 2005.
 The object inside the box is a very blue
 star  within $\simeq$5'' from the Chandra position, with B$\simeq$22.7.
No star with B$<$23.5 is closer to the Chandra positions in the images
of the same night. The image center and size are chosen to allow
 easy identification of the field and serve as a finding chart.
}
\end{figure}
\begin{figure}
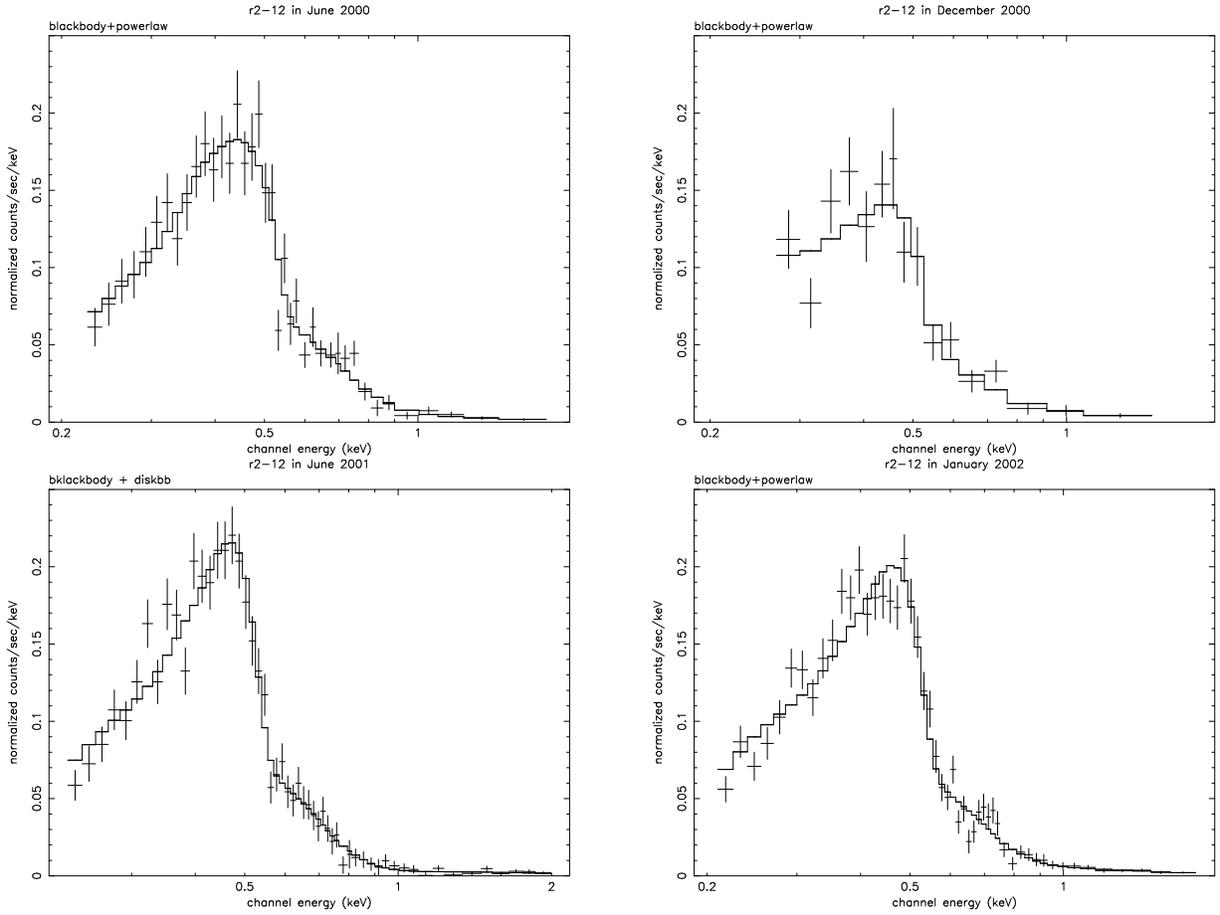

\begin{center}
\includegraphics[width=6cm,angle=-90]{f6a.ps}
\hspace{0.8cm}
\includegraphics[width=6cm,angle=-90]{f6b.ps}
\vspace{1.5cm}
\includegraphics[width=6cm,angle=-90]{f6c.ps}
\hspace{0.8cm}
\includegraphics[width=6cm,angle=-90]{f6d.ps}
\end{center}
\caption{The MOS spectra of source r2-12 in the  bulge exposures,
 and best fit with a blackbody and powerlaw or `diskbb'' model (see Table 6).
The best fit T$_{\rm BB}$ is 60-65 eV in all observations, and
the luminosity appears to have increased in 2001-2002.}
\end{figure}
\begin{figure}
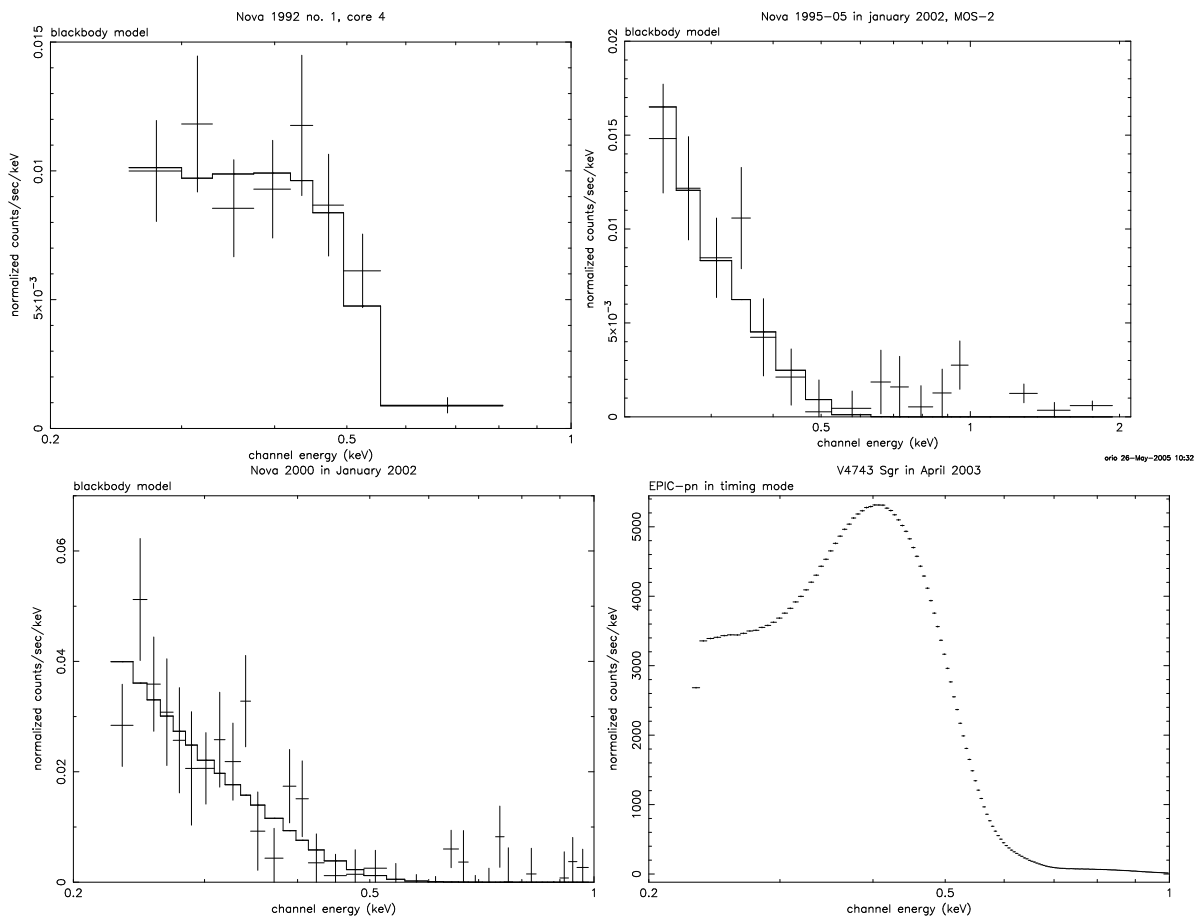

\begin{center}
\includegraphics[width=6cm,angle=-90]{f7a.ps}
\includegraphics[width=6cm,angle=-90]{f7b.ps}
\includegraphics[width=6cm,angle=-90]{f7c.ps}
\includegraphics[width=6cm,angle=-90]{f7d.ps}
\end{center}
\caption{The pn spectra of 3 of the 5 novae in M31, and comparison 
 with the pn spectrum in timing mode (not piled-up) of the Galactic nova
V4743 Sgr, the brightest nova ever
observed in X-rays, 6 months after the outburst. The best fit
blackbody temperatures of the M31 novae are 54 eV for N 1991-1, 34 eV 
for N 1995-5, 49 eV for N 2000. Fits with WD atmospheric models indicate
a temperature around 80 eV for V4743 Sgr. }
\end{figure}
\begin{figure}
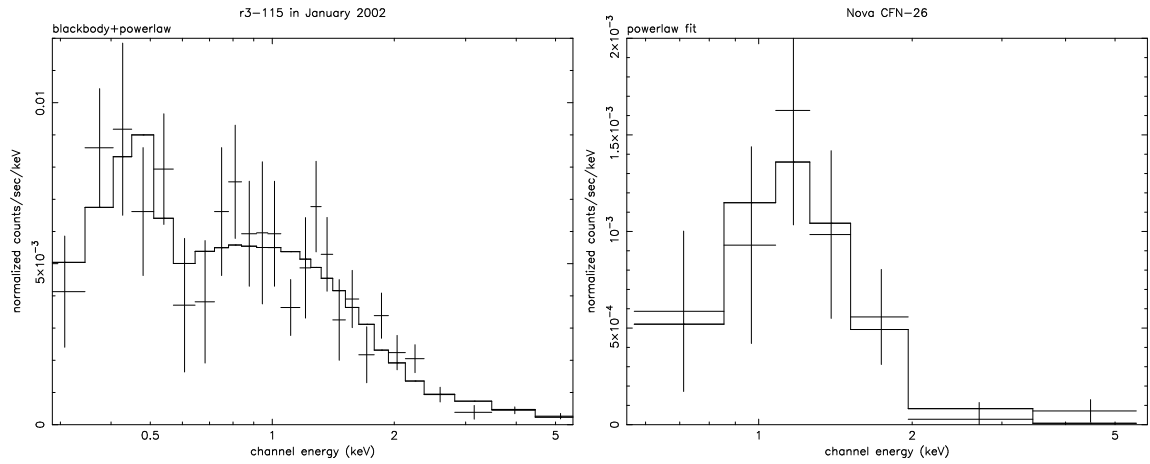

\begin{center}
\includegraphics[width=6cm,angle=-90]{f8a.ps}
\vspace{1.5cm}
\includegraphics[width=6cm,angle=-90]{f8b.ps}
\caption{The panel on the right shows the
spectrum of source r3-115 observed in January 2002. In its 
position, there is no detection above 3 $\sigma$ in the previous XMM-Newton
 exposures, but a supersoft source appeared in the Chandra ACIS-S observation
of October 2001. The source coincides spatially
 with an optically variable object that is thought to possibly be a
 symbiotic nova. The panel on the left shows the spectrum of
recurrent nova Rosino 140 or CFN-26, observed with MOS-2 in the
core-4 exposure.}  
\end{center}
\end{figure}
\clearpage
\begin{deluxetable}{rrrrrrrrr}
\tabletypesize{\scriptsize}
\tablecolumns{9}
\tablewidth{0pc}
\tablecaption{{Exposures of M31 done with XMM-Newton, filter used, nominal
 length of exposure in kiloseconds and interval of time selected
 for the data analysis (effective exposure), date of observation,
 number of Chandra and ROSAT  SSS detected over total number observed,
same ratio for Chandra QSS, and total number of new  SSS discovered in the image.}} 
\tablehead{
\colhead{obs. ID}  & \colhead{region}  & \colhead{filter}   & \colhead{exposure} & \colhead{effective} &
\colhead{date}     & \colhead{old  SSS} & \colhead{old QSS} & \colhead{new SSS }  \\
\colhead{}         & \colhead{}         & \colhead{}        & \colhead{time (ksec)} & \colhead{time (ksec)} &
\colhead{}         & \colhead{}         &   \colhead{}      &  \colhead{}               } 
\startdata
0109270701  & North1 & M  &  58.2 & 54.8 &  2002-1-5  & 0/1  &      & 1   \\           
0109270301  & North2 & M  & 57.2  & 24.3 &  2002-1-26 & 1/3  &      &       \\
0109270401  & North3 & M  & 64.3  & 53.6 &  2002-6-29 & 0/2  &      &      \\
0151580401  & Halo4  &    & 11.4  & 11.4 &  2003-2-6  &   &  0/1    &       \\          
0112570401  & Core1  & M  & 46.0  & 28.0 &  2000-6-25 & 5/14 & 4/5  &   2    \\
0112570601  & Core2  & M  & 13.3  &  9.9 &  2000-12-28 & 4/13 & 4/5  & 1     \\        
0109270101  & Core3  & M  & 57.9  &  33.3 & 2001-6-29  & 6/12 &    & 3  \\ 
0112570101  & Core4  & T  & 64.3  &  61.1 & 2002-1-6   & 8/12 & 4/5    & 4   \\
0112570201  & South1 & T  & 67.3  &  54.4 & 2002-1-12  & 3/5  &   1/1  & 2  \\
0112570301  & South2 &    & 62.8  &  46.7 & 2002-1-24  & 0/2  &        &       \\ 
\enddata
\end{deluxetable}
\clearpage
\begin{deluxetable}{rrrrrrrrrrrr}
\rotate
\tabletypesize{\tiny}
\tablecolumns{11}
\tablewidth{0pc}
\tablecaption{\tiny{SSS detected in the XMM-Newton exposures
of M31, in order of decreasing declination,
and upper limit for Chandra source r2-66. The Chandra  name 
for  SSS from the Di Stefano et al. (2004)
article is in boldface. ``GS'' means presence of  a GALEX UV source in the error circle. 
The columns give: coordinates, name of exposure, identification
with known objects and reference, count rates
in the 02.-2 keV range and 1 $\sigma$e errors, parameters of the best blackbody or powerlaw 
best fit, including either the bolometric luminosity
obtained from a blackbody fit, or the unabsorbed luminosity
in the 0.2-2 keV range obtained from a powerlaw fit. All the models
are obtained with $\chi^2<$1.7, but if 
 $\chi^2 >1.1$, it is noted and discussed in the text. The catalog
name of the Chandra  SSS (from the list of Di Stefano et al.
 2004) is written in boldface. N(H) is written 
 in boldface if the best fit was done with a fixed
 value of this parameter.}}
\tablehead{
\colhead{N} &
 \colhead{$\alpha$}  & \colhead{$\delta$} & \colhead{exp.} & ID/ref. & \colhead{pn}  &
\colhead{MOS-2} & \colhead{N(H)} & \colhead{T(bb)} & \colhead{$\nu$} & 
 \colhead{L $\times 10^{36}$} \\
\colhead{}    & \colhead{(2000)} & \colhead{(2000)}   & \colhead{} & \colhead{}&
\colhead{10$^{-3}$cts s$^{-1}$} & \colhead{10$^{-3}$cts s$^{-1}$} & \colhead{10$^{21}$ cm$^{-2}$} &  \colhead{(eV)}  &
\colhead{}  & \colhead{erg s$^{-1}$}}
\startdata
1 & 00 42 16.24 & 40 48 03.9  & S-1 & GS, OB ass. (1) & 1.48$\pm$0.27 &    & 1.4  & 122   &  & 1.12 \\ 
  &             &             &     & RX J004217.4+404812 (2)     &    &      &      && \\
     & & & & & & & & & & & \\
2     & 00 42 12.8 & 41 05 58.9  & C-3 & {\bf r3-122}  & 1.25$\pm$0.48 &    &      &        & &  \\
      &            &             &     & B=18.7, R=18.4 &             &    &      &        & &  \\
 & & & & & & & & & &  \\
3 & 00 41 54.08 & 41 07 23.3  & C-2 & N 92-01 (3)   & $\leq$4.30     &  &  &  &   \\
  &             &             & C-4 &               &                & 3.05$\pm$0.25  & 0.8 & 54  & & 786 \\ 
  &             &             & S-1 &               & 5.72$\pm$0.56  & 0.51$\pm$0.10  & 1.9  & 51   & & 162 \\
     & & & & & & & & & &  \\
4 & 00 43 08.60 & 41 07 30.6 & C-1 &  r3-131, SG,V$\simeq$18 & 
                                                1.24$\pm$0.54 &     & 1.7   & 138  & & 1.4 \\
  &             &            & C-2 &  GS        & $\leq$5.65    &      &    &    & &   \\
  &             &            & C-3 &          & 1.31$\pm$0.38 &      &    &    & &   \\
  &             &            & C-4 &          & 1.30$\pm$0.43 &      & 1.0   & 153    & &  \\
     & & & & & & & & & &  \\
5  & 00 42 42.13 & 41 12 20.1  & C-4 & GS, N-1995 (3) & 2.31$\pm$0.80 &         &    & & &  \\ 
     & & & & & & & & & &  \\
6  & 00 42 47.0 & 41 14 12.4   & C-1 &  {\bf r2-65} & $\leq$7.15    &                &        & & & \\
   &            &              & C-2 &              & 4.27$\pm$2.05 &      &    & & & \\
   &            &              & C-4 &              & 3.49$\pm$0.74 &                & {\bf 1} & 50 & & 20.2 \\
    & & & & & & & & & &  \\
7 & 00 42 52.4  & 41 15 39.7  & C-1 & {\bf r2-12}       &               & 60.48$\pm$1.57 & 1.1 & 68  & 2.9 & 417 \\
  &             &              & C-2 &             &                     & 47.10$\pm$2.30 & 0.8 & 63   & 2.5 & 337 \\
  &             &              & C-3 &             &                    & 64.05$\pm$1.34 & 1.8 & 60   & &  2484 \\
 &             &              & C-4 &             &                      & 63.35$\pm$1.11 & 1.7 & 60 & 2.5 & 1178 \\
    & & & & & & & & & &  \\
8 & 00 42 50.4  & 41 15 56.2  & C-1 & {\bf r2-56}       &           & 1.73$\pm$0.79   &  & &   &   \\
  &             &              & C-2 &                   &               & 4.07$\pm$0.80      &  &  & & &  \\   
  &             &              & C-3 &                   &               & 5.86$\pm$0.50      &  &  & & &  \\
  &             &              & C-4 &                   &               & 3.97$\pm$0.55     &  & &  & &  \\
    & & & & & & & & & &  \\
9 & 00 42 59.2 & 41 16 44.3 & C-1 & N95-05/{\bf r2-63} & $\leq$1.02              & & & & & \\
    &            &             & C-2 &     (3)         & $\leq$4.64           & $\leq$0.37 & & & & \\
    &            &             & C-3 &                    &  8.81$\pm$1.35  &               & & & & \\
    &            &             & C-4 &                    &                 & 2.82$\pm$0.39 & {\bf 0.8} & 34  & & 364 \\
    & & & & & & & & & &  \\
10 & 00 42 43.90 & 41 17 55.5  & C-1 & {\bf r2-60}/N-2000 (3) & $\leq$1.36   &                & & & &  \\
  &             &              & C-2 &               & $\leq$8.07   &              & & & &   \\
  &             &              & C-3 &               & 4.22$\pm$1.28 & 1.01$\pm$0.41 & & & &   \\
 &             &              & C-4 &               & 4.34$\pm$0.74 & 0.97$\pm$0.30 & {\bf 0.7} & 49 & & 20  \\
    & & & & & & & & & &  \\
11  & 00 43 19.4 & 41 17 59.0       & C-1 & GS, (4) &   & 30.221$\pm$1.11 & 0.81 & 65 & & 295  \\
    &            &                  & C-2 &             &   $\leq$2.28        &      &    & & & \\
    &            &                  & C-3     &         & $\leq$0.97          &      &    & & & \\
    &            &                  & C-4     &         & $\leq$0.71          &      &    & & & \\
    & & & & & & & & & &  \\
12 & 00 43 08.30 & 41 18 20.0     & C-1 &  (5)  & $\leq$0.98     &   $\leq$0.66          & & & & \\
    &             &                 & C-2 &             & $\leq$2.60     &                & & & & \\
    &             &                 & C-3 &             & 10.68$\pm$1.01 & 1.73$\pm$0.35 & 0.7f & 34 & & 300 \\
    &             &                 & C-4 &             & $\leq$0.84    &                & & & & \\ 
    & & & & & & & & & & \\
13  & 00 42 49.0 & 41 19 47.1       & C-1 & {\bf r2-66} & $\leq$1.00         &       & & & &   \\
 &            &                  & C-2 &    GS       &         &  $\leq$0.37          & & &   \\
 &            &                  & C-3 &             &         &  $\leq$0.82          & & &   \\
  &            &                  & C-4 &             &                     & $\leq$0.02 & & &   \\
    & & & & & & & & & &  \\
14  & 00 43 18.85 & 41 20 18.5     & C-1 & r3-8 & $\geq$75.14 & 16.64$\pm$0.84 & 2.5 & 64  & & 653 \\
    &             &                 & C-2 & RX J004318.6+412024 (2) & 12.30$\pm$2.78 & & 1.4 & 73 & & 31 \\
    &             &                 & C-3 & GS            & 6.95$\pm$1.10  &   & {\bf 0.7} & 48 & & 28 \\
    &             &                 & C-4 &            & 47.16$\pm$1.26     & 9.43$\pm$0.46  & 2.3  & 64 & & 367 \\
    & & & & & & & & & & \\
15 & 00 42 55.34 & 41 20 44.7 & C-1 & N96-05 (3)   & $\leq$1.77    &               &  & &  &\\
   &             &             & C-2 &             & $\leq$2.21    &               & &  & & \\
   &             &             & C-3 &             & $\leq$1.02   &               & &  & & \\
   &             &             & C-4 &             & 3.49$\pm$0.70 &                   & 0.85  & 39  & & 155 \\
\enddata
\tablecomments{References: 1) Hill et al. 1995, 2) Kahabka 1999, 3) Shafter \& Irby 2001,
 4) Osborne et al., 2001, 5) Shirey, 2001.}
\end{deluxetable}
\clearpage
\begin{deluxetable}{rrrrrrrrrrrrr}
\tabletypesize{\tiny}
\rotate
\tablecolumns{13}
\tablewidth{0pc}
\tablecaption{\small{Same entries as in Table 2, for the Quasi Soft Sources
(QSS) detected with Chandra and/or with XMM-Newton. The count rates
are measured in the 0.2-10 keV range (total), and  the S, M and H ranges are
 defined in the text. The catalog
name of the Chandra QSS (from Di Stefano et al. 2004) is written in 
boldface.}}
\tablehead{
\colhead{N} &
 \colhead{$\alpha$}  & \colhead{$\delta$} & \colhead{exp.} & \colhead{ID/ref.} &  \colhead{total}  & 
\colhead{S} & \colhead{M} & \colhead{H} & \colhead{N(H)} & \colhead{T(bb)} &  \colhead{$\nu$} &
 \colhead{L $\times 10^{36}$} \\
\colhead{}    & \colhead{(2000)} & \colhead{(2000)}   & \colhead{} & \colhead{} & 
\colhead{10$^{-3}$cts s$^{-1}$} & \colhead{10$^{-3}$cts s$^{-1}$} & \colhead{10$^{-3}$cts s$^{-1}$} & \colhead{10$^{-3}$cts s$^{-1}$} &
 \colhead{10$^{21}$ cm$^{-2}$} &  \colhead{(eV)}  &
\colhead{}  & \colhead{erg s$^{-1}$}}
\startdata
 1 & 00 39 57.78 & 40 27 23.7 &   S-2/pn & GS & 8.03$\pm$0.84 & 7.09$\pm$0.57 & 1.23$\pm$0.31 & 
 & 1.7 & 156  & & 5.4 \\
   &             &            & S-2/MOS2   &    & 1.76$\pm$0.45 & 1.01$\pm$0.28 & 0.07$\pm$0.11 & 0.46$\pm$0.21 & & & & \\
    & & & & & & & & & &  & & \\
 2 & 00 41 08.34 & 40 51 29.0 &   S-1/pn & & 1.76$\pm$0.45  & 1.01$\pm$0.28 & 0.07$\pm$0.11 &  0.46$\pm$0.21
   & 2.4  & 80    &  & 3.3  \\ 
    & & & & & & & & & &  & & \\
 3 & 00 41 36.5 & 41 00 17.0 &   S-1/pn & {\bf s1-41} & 1.51$\pm$0.47 & 0.96$\pm$0.31   & 0.96$\pm$0.31
 & 0.11$\pm$0.23 & 3.6   & 155  & &  2.3 \\ 
    & & & & & & & & & &  & & \\
 4 & 00 42 07.91 & 41 04 34.2 & C-1/pn   & RX J004208.2  & 5.37$\pm$0.64 & 2.71$\pm$0.43 &
1.19$\pm$0.27 &  0.18$\pm$0.20 & 0.9      &  &  2.1  & 7.6 \\
   &             &            & C-2/pn   & +410438? (1) & 2.68$\pm$0.86 &  &
  & & 0.8  &  & 2.3     & 10 \\
   &             &            & C-3/pn   &         & 7.07$\pm$0.76 &  4.25$\pm$0.50 & 2.16$\pm$0.32 &
 0.47$\pm$0.39      & 0.9           & & 2.3     & 9.2 \\
   &             &            & C-4/MOS2   &     &  1.92$\pm$0.22   & 0.42$\pm$0.21 & 1.61$\pm$0.32 &
0.44$\pm$0.14 & {\bf 0.9} & & 1.6 &    8.9 \\
    & & & & & & & & & & & &   \\
 5 & 00 42 28.67 & 41 04 36.3 & C-1/pn   & r3-111 (2) & 41.57$\pm$1.87 & 26.36$\pm$1.13 & 10.75$\pm$0.85 &
 3.60$\pm$0.82    &  1.4    &   & 2.8  & 40  \\
   &             &            & C-2/pn   &        & 54.78$\pm$2.40 &  29.63$\pm$2.1 & 14.37$\pm$1.47 &
 5.56$\pm$1.08    &  {\bf 1.4} &   &  2.7    & 63     \\
   &             &            & C-3/pn   &     & 51.78$\pm$1.62 & 29.67$\pm$1.17 & 12.06$\pm$0.74 & 
6.81$\pm$0.67     &  {\bf 1.4} & & 2.6  & 46     \\
   &             &            & C-4/pn   &        & 55.93$\pm$1.62 &   29.26$\pm$0.99 & 16.56$\pm$0.68
 & 8.71$\pm$0.59   &  1.4    &  &  2.3  &  54 \\ 
    & & & & & & & & & &  & & \\
 7 & 00 42 36.5  & 41 13 50.1  & C-1/MOS2   & {\bf r2-42} &   1.04$\pm$0.37 &  0.58$\pm$0.29 & 0.08$\pm$0.17 &
 & &  & & \\
   &             &            &  C-2/MOS2   &         &   2.64$\pm$0.88 &  1.18$\pm$0.69 & 0.91$\pm$0.45 & &
  & & &  \\
   &             &            & C-3/MOS2  &   & 3.31$\pm$0.50  &  2.10$\pm$0.41 & 0.87$\pm$0.25 & 0.24$\pm$0.19
& &  & & \\
   &             &            & C-4/MOS2  &   & 2.10$\pm$0.41  &                &                &  &   & &  & \\
    & & & & & & & & & &  & & \\
 8  & 00 42 44.3  & 41 16 07.3 & C-1/pn   & {\bf r1-9}   & $\leq$39.91 & & & &   1.3 & & 2.7 & 13 \\
    &             &            & C-2/pn   &  PN (3)      & $\leq$39.75 & & & &   0.7 & & 2.4 & 66 \\
    &             &            & C-3/pn   &              & $\leq$46.24  & & & & & & & \\
    &             &            & C-4/pn   &            & $\leq$188.40 & & &  & 0.72 & & 2.1 & 68.6 \\   
    & & & & & & & & & &  & & \\
 9 & 00 42 39.2 & 41 14 24.4 & C-1/MOS2 & {\bf r2-62} &   7.47$\pm$0.69 &  2.84$\pm$0.45 & 2.45$\pm$0.36 & 1.11$\pm$0.22 &
 1 & & 2.1 & 10.6 \\
      &            &            & C-2/MOS2 &          &   7.32$\pm$0.67 & 3.17$\pm$0.50 & 2.61$\pm$0.35 &
 0.50$\pm$0.16 & 1.0 & & 2.5 & 14 \\
      &            &            & C-3/MOS2 &          &   4.54$\pm$0.53 & 2.39$\pm$0.49 & 1.19$\pm$0.12 &
 0.40$\pm$0.16 &  1.0    & & 2.6 & 8.2 \\
     &            &            & C-4/MOS2 &             & 8.94$\pm$0.56   &  4.01$\pm$0.43 & 2.62$\pm$0.27 &
1.56$\pm$0.19 & 1.08   &  & 2.3 & 10.7  \\
    & & & & & & & &  & & & &  \\
 10 & 00 42 38.6  & 41 15 26.3 & C-1/pn & r2-54    & $\leq$1.50              &                & & & & \\
    &             &            & C-2/pn &   &     $\leq$8.36  & &  & & &  \\
    &             &            & C-3/MOS2 & & 3.96$\pm$0.65 & 3.04$\pm$0.60 & 0.69$\pm$0.24 & &  1          & & 2.0 & 6.2 \\
    &             &            & C-4/MOS2 &     & 3.47$\pm$0.47 & 2.82$\pm$0.44 & 1.86$\pm$0.38 & 0.48$\pm$0.15&         1 & &  2.4  & 6.3 \\
    & & & & & & & & & &  & & \\
 11 & 00 43 14.3 & 41 16 50.1 & C-1/pn & {\bf r3-11} & 1.97$\pm$0.98 & 1.21$\pm$0.82   & 0.70$\pm$0.30  & 0.28$\pm$0.22 &        & &     &      \\ 
    &            &            & C-2/pn &             & $\leq$2.33       & &  &  & &     & &      \\
    &            &            & C-3/pn &         & $\leq$2.15  &  & &             &  & &     &      \\
    &            &            & C-4/pn &          & 3.72$\pm$0.98 &  2.32$\pm$0.80 & &  0.72$\pm$0.35 &      &     & &      \\
    & & & & & & & & & &  & & \\
12 & 00 43 06.9 & 41 18 09.0 & C-1/pn & {\bf r3-115} & $\leq$1.0      && &                  &     & & & \\
   &            &             & C-2/pn &              & $\leq$3.24 &     & & &     & & & \\
   &            &             & C-3/pn &              & $\leq$1.10 &    & &  &     & & & \\
   &            &             & C-4/pn &              & 8.29$\pm$0.42    &  4.05$\pm$0.30  & 
 2.23$\pm$0.21 & 1.41$\pm$0.17   &    1.16 & 69 & 1.9 & 13.3 \\
    & & & & & & & & &  & & &  \\
 13 & 00 45 58.1 & 41 35 02.2   & H-4/pn & {\bf n1-31} &  $\leq$4.30  & &                &       & &     &      \\
  & & & & & & & & & & & & \\
 14  & 00 46 49.27 & 42 09 26.2   &   N-3/pn &  & 1.33$\pm$0.28 & &    & 0.10$\pm$0.11 &
          10 & 117 & & 1.1 \\
& &    & & & & & & & & & & \\
 15  & 00 46 16.7 & 41 36 56.0    & H-4/pn &   {\bf n1-8} & $\leq$4.30  &   & &     & &     &      \\
    & & & & & & & & & &  & & \\
 16  & 00 46 23.81 & 42 21 33.7   &   N-3/pn & & 1.79$\pm$0.45 & 1.41$\pm$0.34 &  0.17$\pm$0.11 &
 0.19$\pm$0.25 & 1.5 & 123 & &  0.6 \\
    & & & & & & & & & &  & & \\
 17  & 00 47 00.71 & 42 21 51.9   &   N-3/pn & & 3.19$\pm$0.45 &  & 0.30$\pm$0.15 & 
0.26$\pm$0.15        & 3.1 & 129 & & 3.0 \\ 
    & & & & & & & & & &   & & \\
 18  & 00 46 03.00 & 42 24 31.8  &  N-3/pn  & B$\simeq$15 & 2.24$\pm$0.51 & 1.31$\pm$0.34 &  0.05$\pm$0.30 & 0.50$\pm$0.23
    & 1.3  & 137 & & 1.61 \\
\enddata
\tablecomments{References: 1) Supper et al. 1997, 2) Kong et al. 2002, 3) Ciardullo et al. 2002}
\end{deluxetable}
\clearpage
\begin{deluxetable}{rrrrrrrrrrrrr}
\tabletypesize{\tiny}
\rotate
\tablecolumns{12}
\tablewidth{0pc}
\tablecaption{\small {Supernova remnants with hardness ratios of SSS or QSS,
detected with Chandra (above the line) and with XMM-Newton (below
 the line). The three sources above the
line were also detected with XMM-Newton, but the
 source confusion near the bulge prevents obtaining reliable count rates.
Only EPIC-pn count rates are given. The ``total count rate''
is in the 0.2-10 keV range, count rates S, M and H are defined in the text.
The best blackbody fit parameters are given in the last three columns.}}
\tablehead{
\colhead{N} &
 \colhead{$\alpha$}  & \colhead{$\delta$} & \colhead{exp.} & \colhead{ID} & \colhead{ref.} &
 \colhead{\rm total} & \colhead{S}   & \colhead{M} & \colhead{H} &
\colhead{N(H)} & \colhead{T(bb)} & \colhead{L $\times 10^{36}$} \\ 
\colhead{}    & \colhead{(2000)} & \colhead{(2000)}   & \colhead{} & \colhead{} &
 \colhead{} & 
\colhead{10$^{-3}$ s$^{-1}$} & \colhead{10$^{-3}$ s$^{-1}$}& 
\colhead{10$^{-3}$ s$^{-1}$} & \colhead{10$^{-3}$ s$^{-1}$} &
 \colhead{10$^{21}$ cm$^{-2}$} &  \colhead{(eV)}  &
\colhead{erg s$^{-1}$} }
\startdata
 1  & 00 41 35.6  & 41 06 56.8 & S-1 &  SSS {\bf s1-42}  & 1,2,3 & $\geq$1.86   & & & & & & \\
    &             &            &     & RX J004135.8+4110657 &    &          & & & & & &  \\
& & & & & & & & & & & &   \\
 2  & 00 42 43.0 & 41 16 03.1 &   &  SSS {\bf r1-35}, S And    & 1     &    & & & & & & \\
& & & & & & & & & & & &  \\
 3 &   00 42 49.0 & 41 19 47.1 & C-1 & {\bf QSS r2-66} & 2 & $\leq$1.11    &   &      &  & &     &      \\
   &              &             & C-2 &  B-96      &           & $\leq$0.82  &  & & & & &      \\
   &              &             & C-3 &        & & $\leq$1.62     &           &  & &  &  &      \\
   &              &             & C-4 &             &             &   $\leq$0.42 &  & & &     & &      \\
 & & & & & & & & & & &  & \\
\hline
& & & & & & & & & & &   \\
 4  & 00 43 43.80 & 41 12 32.4 & C-1   & RX J004344.1+411219 & 1,2,4,5 & 4.85$\pm$0.75  & 4.26$\pm$0.55 & 0.41$\pm$20 &  & 2.8  & 144 & 7.5  \\
 & & & & QSS & & & & & & & & \\
  &             &            & C-2   & radio s. & & $\geq$2.68 &         &     &     &     & & \\
 &             &            & C-3   &  SSS & & 4.52$\pm$0.95  & 4.06$\pm$0.33 &
0.30$\pm$0.33 & 0.28$\pm$0.43 & 4.3 & 119  & 18.1 \\
  &             &            & C-4   &  B-166  & & 5.46$\pm$0.65 & 4.89$\pm$0.48
& 0.43$\pm$0.22 & &  3.0 & 140  & 10.8  \\
 & & & & & & & & & & & \\ 
 5  & 00 43 27.78 & +41 18 29.2 & C-1 & r3-63 & 1,6 & 27.91$\pm$1.52  &
 27.36$\pm$1.32 & 0.80$\pm$0.36 & 0.0$\pm$0.41 & & & \\
    &        &     & C-3 & B-142 &   & 26.20$\pm$1.13 & 25.03$\pm$1.15 &
 0.42$\pm$0.24 & 0.34$\pm$0.33 & 1.3 & 129 & 16.3  \\
    &         &      & C-4 &  SSS &   & 27.20$\pm$0.95 & 23.77$\pm$0.83
 & 1.51$\pm$0.26 & 0.81$\pm$0.25 & 1.4 & 133 & 18.0 \\
 & & & & & & & & & & & &  \\ 
 6  & 00 42 48.97 & 41 24 06.9 &  C-2 & B-95,r3-84 & 1,7 & 3.16$\pm$1.0  & 3.11$\pm$0.77 &   &  & & &  \\
    &             &            & C-3 &  SSS & & 3.11$\pm$0.77   & 2.91$\pm$0.50 &
0.02$\pm$0.20  &  & & & \\
    &             &            &  C-4 & & & 3.92$\pm$0.67  & 3.00$\pm$0.50 & 0.64$\pm$0.27 & & 1. & 126 & 2.4  \\ 
  & & & & & & & & & & & & \\
 7 & 00 43 39.22 & 41 26 55.3  & C-2 &  B-158 &2,3,9,10 & 6.83$\pm$1.34 & 
6.81$\pm$1.21 &  & & {\bf 1.3} & 148 & 6.0 \\
    &            &             & C-4 &   SSS & & 5.33$\pm$0.66 &  4.36$\pm$0.41 &
 0.08$\pm$0.17 & 0.51$\pm$0.21 & {\bf 1.3}  & 148 &  5.3 \\
    &            &             & N-1 &  &  & 8.11$\pm$0.61 & 8.00$\pm$0.54
 & 0.19$\pm$0.15 &  0.16$\pm$0.19  & 1.3 & 123 & 6.9 \\
    & & & & & & & & & & &  \\
 8 & 00 44 51.00 & 41 29 05.8 & N-1 & WB92a, QSS & 5 & 3.77$\pm$0.66 & & & & 2.7 &
 116 &  9.7 \\
    & & & & & & & & & & & \\
 9  & 00 45 14.35 & +41 36 16.9 & N-1 & B-90,r3-67, QSS & 1,9 & 2.84$\pm$0.57 & 2.67$\pm$0.40
 & 0.25$\pm$0.15 &  & 4.2  & 123  & 17.5 \\
    & & & & & & & & & & & \\
 10  & 00 46 27.75 & 42 08 00.3 & N-3 & B-476, QSS & 2 &2.47$\pm$0.51 & 1.40$\pm$0.36  &
0.40$\pm$0.15 & 0.15$\pm$0.20 & 2.1 & 123  & 2.9 \\
\enddata
\tablecomments{References: 1) Di Stefano et al. 2004, 2) Braun 1995, 3) Supper et al. 1997, 
 4) Bystedt et al. 1984, 5) Walterbros \& Braun 1992, 6) Kong et al.,
 2002, 7) Kong et al. 2003,  8) Williams et al. 2004a, 9) Williams et al. 2004b,
 10) Roth et al. 2004, 11) Kahabka 1999.}  
\end{deluxetable}
\clearpage
\begin{deluxetable}{rrrr}
\tabletypesize{\scriptsize}
\tablecolumns{4}
\tablewidth{0pc}
\tablecaption{ \small {2 $\sigma$ range of the parameters for the blackbody
fit to the observations of source r3-8.}}
\tablehead{
\colhead{observation} & \colhead{N(H)} & \colhead{T(bb)} &  \colhead{L $\times 10^{38}$} \\
\colhead{}            & \colhead{10$^{21}$ cm$^{-2}$}     & \colhead{eV}      &  \colhead{erg s$^{-1}$}}\startdata
Core-1, pn & 3.8-4.7 & 52-60 & $\leq$46.20 \\
 & & & \\ 
Core-1, MOS & 4.1-5   & 52-55 & $\leq$11.55 \\
 & & & \\
Core-2, pn & $\leq$3.7 & 32-130 & $\leq$2.89 \\
 & & & \\
Core-3, pn & $\leq$0.9 & 35-79 & $\leq$4.04 \\
 & & & \\
Core-4, pn & 6.6-7.6 & 44-52 & 19.47-30. \\
 & & & \\
Core-4, MOS & 2.1-3.7 & 53-66 & $\leq$33 \\ 
 & & & \\
\enddata
\end{deluxetable} 
\clearpage
\begin{deluxetable}{rrrrrr}
\tabletypesize{\scriptsize}
\tablecolumns{6}
\tablewidth{0pc}
\tablecaption{ \small {2 $\sigma$ range of the parameters for the blackbody
and powerlaw fit to the observations of source r2-12.}}
\tablehead{
\colhead{observation} & \colhead{N(H)} & \colhead{T(bb)} &  \colhead{L$_{\rm bol} \times 10^{38}$}
& \colhead{$\nu$} & \colhead{T$_{\rm d,in}$} \\
\colhead{}            & \colhead{10$^{21}$ cm$^{-2}$}     & \colhead{eV}      &  \colhead{erg s$^{-1}$} &
\colhead{} & \colhead{keV} }
\startdata
Core-1 & 0.83-1.26   & 65-75 & 2.54-5.70 & 0.26-4.12 & \\
 & & & & \\
Core-2 & 0.27-1.18 & 40-78 & 1.53-10.22 & $\leq$4.4 & \\
 & & & & \\
Core-3 & 1.63-2.04 & 58-63 & 10.51-24.07 &  & $\simeq$1.57 \\ 
 & & & & \\
Core-4 & 1.65-1.93 & 57-63 & 10.40-20.22 & 1.8-3.4 & \\
 & & & & \\
\enddata
\end{deluxetable}

\end{document}